\documentclass{egpubl}
\usepackage{eurovis2026}

\usepackage[T1]{fontenc}
\usepackage{dfadobe}  

\biberVersion
\BibtexOrBiblatex
\usepackage[backend=biber,bibstyle=EG,citestyle=alphabetic,backref=true,maxbibnames=99]{biblatex} 
\electronicVersion
\PrintedOrElectronic
\ifpdf \usepackage[pdftex]{graphicx} \pdfcompresslevel=9
\else \usepackage[dvips]{graphicx} \fi

\usepackage{egweblnk}

\newcommand{\mypar}[1]{\smallskip\noindent\textbf{#1}}

\title[BlockSets]%
      {BlockSets: A Structured Visualization for Sets with Large Elements} 
      
\makeatletter
\renewcommand*{\@fnsymbol}[1]{\ensuremath{\ifcase#1\or *\or \dagger\or \ddagger\or
   \mathsection\or \mathparagraph\or \|\or **\or \dagger\dagger
   \or \ddagger\ddagger \else\@ctrerr\fi}}
\makeatother
\author[Neda Novakova, Veselin Todorov, Steven van den Broek, Tim Dwyer, Bettina Speckmann]
{\parbox{\textwidth}{\centering Neda Novakova$^{1}$\thanks{These authors contributed equally.} \orcid{0009-0003-5693-8786}, Veselin Todorov$^{1}$\footnotemark[1]\orcid{0009-0002-9768-253X}, Steven van den Broek$^{1}$\orcid{0009-0005-6677-3916}, Tim Dwyer$^{2}$\orcid{0000-0002-9076-9571}, and Bettina Speckmann$^{1}$\orcid{0000-0002-8514-7858} 
        }
        \\
{\parbox{\textwidth}{\centering $^1$ TU Eindhoven, The Netherlands\\
       $^2$ Monash University, Australia
       }
}
}

\usepackage{amsmath}
\usepackage{xcolor}
\usepackage{stfloats}
\usepackage{subcaption}
\usepackage{booktabs}
\usepackage{multirow}
\usepackage{wrapfig}
\newcommand{\etal}{\emph{et al.}}

\usepackage{rotating}

\graphicspath{{figures/}}

\begin{document}

\teaser{
 \includegraphics[width=\linewidth]{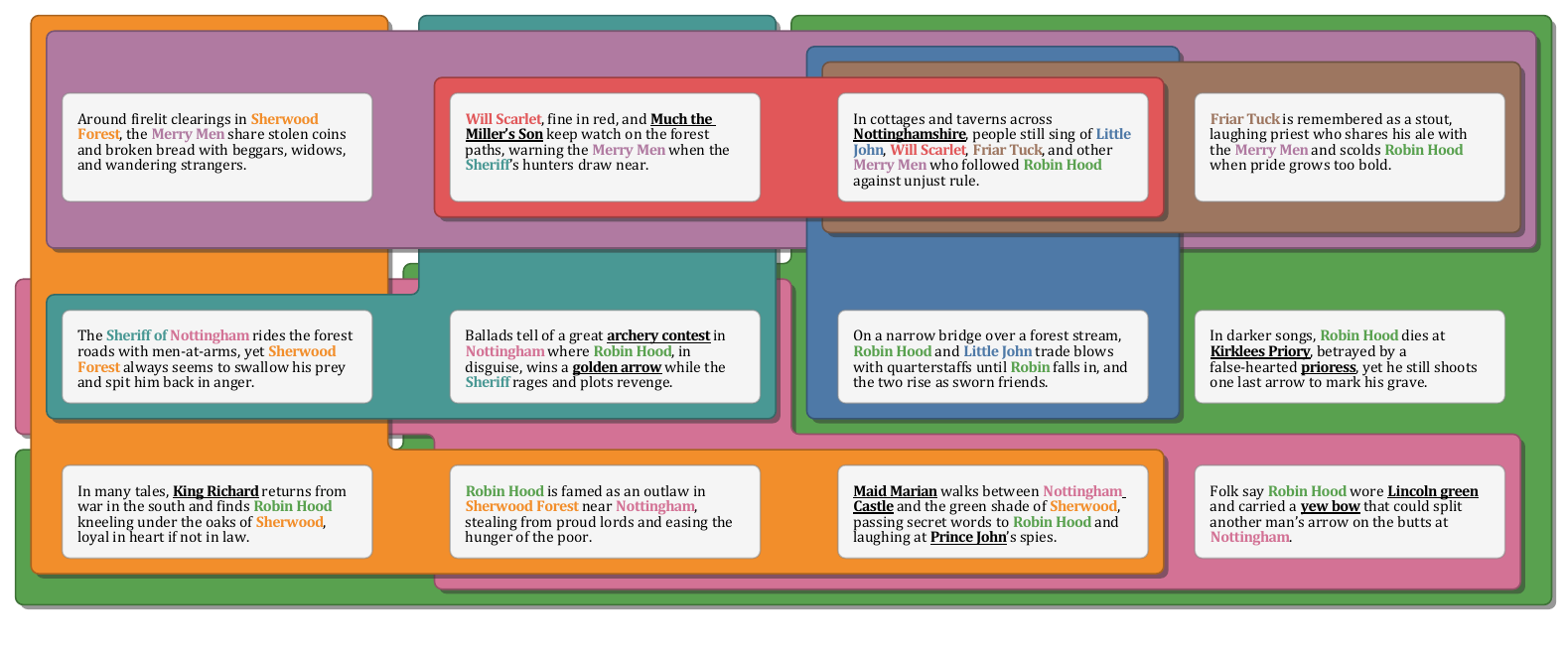}
 \centering
  \caption{Using BlockSets to visualize statements, generated using GPT 5.1, summarizing the tales of Robin Hood.}
\label{fig:teaser}
}

\maketitle
\begin{abstract}
Visualizations of set systems frequently 
use enclosing geometries for the sets in combination with reduced representations of the elements, such as short text labels, small glyphs, or points. 
Hence they are generally unable to adequately represent sets whose elements are larger text fragments, images, or charts. In this paper we introduce \textbf{BlockSets}, a novel set visualization technique specifically designed for sets with large elements.
BlockSets places the elements on a grid and uses rectilinear shapes as enclosing geometries. We describe integer linear programs that find high-quality layouts of the elements on the grid. Since not all set systems allow a compact contiguous representation in this form, we also present an algorithm that splits the visualization into parts when needed; our visual encoding highlights the parts for the user in the final visualization. 
BlockSets utilizes orthoconvex shapes which offer a good trade-off between compactness and readability. Finally, BlockSets renders the enclosing geometries as stacked opaque shapes. We describe an algorithm that finds a stacking order such that all shapes can be inferred. 
Such a stacking does not have to exist, but our algorithm did find a stacking for all real-world data sets that we tested. 

\ccsdesc[500]{Human-centered computing~Visualization techniques}
\ccsdesc[300]{Human-centered computing~Graph drawings}
\ccsdesc[300]{Human-centered computing~Information visualization}

\end{abstract}  
\section{Introduction}
Set visualization is a large and active field of research.
Due to the huge number of possible set systems and the complex ways in which sets can intersect, finding understandable visual representations of those intersections 
is a challenging visualization problem. Furthermore, showing set membership of elements is often not the only challenge: the elements themselves can have associated data attributes that need to be incorporated into the visualization. Set visualizations need to find a balance between the visual requirements of the sets and those of the elements. 
Here, most set visualization techniques favor the sets over the elements. They typically encode set relations using enclosing geometries and allocate only little space to the elements, which are represented by a short text label, a small glyph, a point, or sometimes even not at all. Hence they are generally unable to represent set systems well whose elements are larger text fragments, images, charts, or a combination.

\begin{figure*}[b]
    \centering
    \includegraphics[page=1]{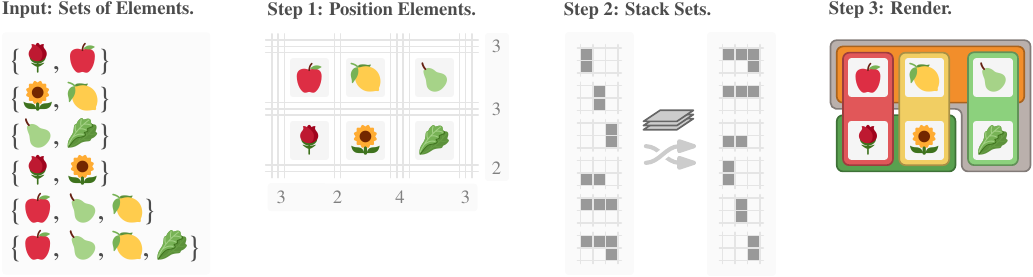}
    \caption{Pipeline for a single component of the visualization. BlockSets receives as input, elements -- each with an associated visual representation -- that are grouped into sets. First, BlockSets assigns the elements to cells in a rectangular grid (detailed in \autoref{sec:layout-algorithm}). The positioning of the elements determines the cells occupied by each set. Appropriate space (see the numbers in the figure) is reserved in the grid for placing the enclosing shapes that represent the sets. In the next step, BlockSets determines a stacking order on the sets such that all shapes can be inferred (\autoref{sec:stacking-algorithm}). Lastly, the enclosing shapes are created, colors are assigned, and the final visualization is rendered.}
    \label{fig:pipeline-single-component}
\end{figure*}

Such data arises, for example, in the form of text summaries produced by large-language models.
Though a text summary is traditionally presented as a sequence of statements, the entities in the text, such as people, objects or places, connect statements in a complex way.
The statements and entities can be thought of as a set system with the entities as the sets and statements as elements of all sets whose entities they mention. 
Such statement-entity data was the initial motivation for our work; we view set visualizations as a promising tool to create structured, compact, non-linear layouts which can help with the comprehension of text.
Another example are small multiples with overlaid set systems, such as a product inventory where the sets are categories and the charts depict the sales over time. Or set systems where the elements are images. 

The examples above suggest several specific requirements for set visualizations with large elements. First of all, the visualization needs to allocate a sizeable, typically rectangular, region to display the element data. Second, the elements should be presented with some form of alignment, to enable easy comparisons for small multiples and support reading comprehension for text summaries. Third, since the elements are large, it is important that they are placed somewhat compactly, to optimally utilize (screen) space. These requirements are, of course, in addition to the standard readability requirements for set visualizations which posit that the user can easily determine set memberships and elements.

\begin{figure*}
    \centering
    \begin{subfigure}{\textwidth}
        \centering
        \includegraphics[width=.9\textwidth]{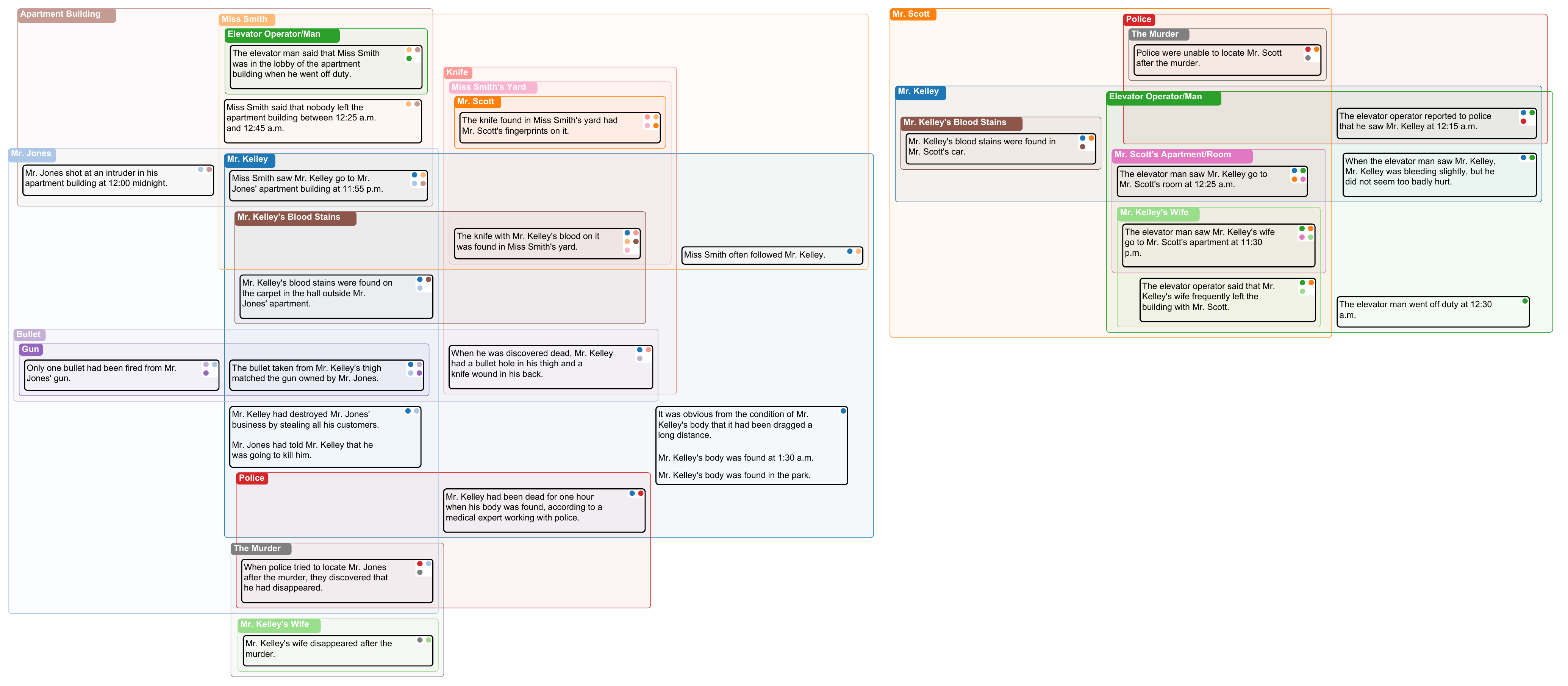}
        \caption{RectEuler~\cite{paetzold2023recteuler}: split into two parts, 7 repeated sets, not compact and no alignment, statements sometimes grouped into items, color blending of transparent sets hinders readability.}
        \label{fig:murder:RectEuler}
    \end{subfigure}

    \vspace{2mm}
    
    \begin{subfigure}[t]{0.5\textwidth}
        \includegraphics[width=1.45\textwidth]{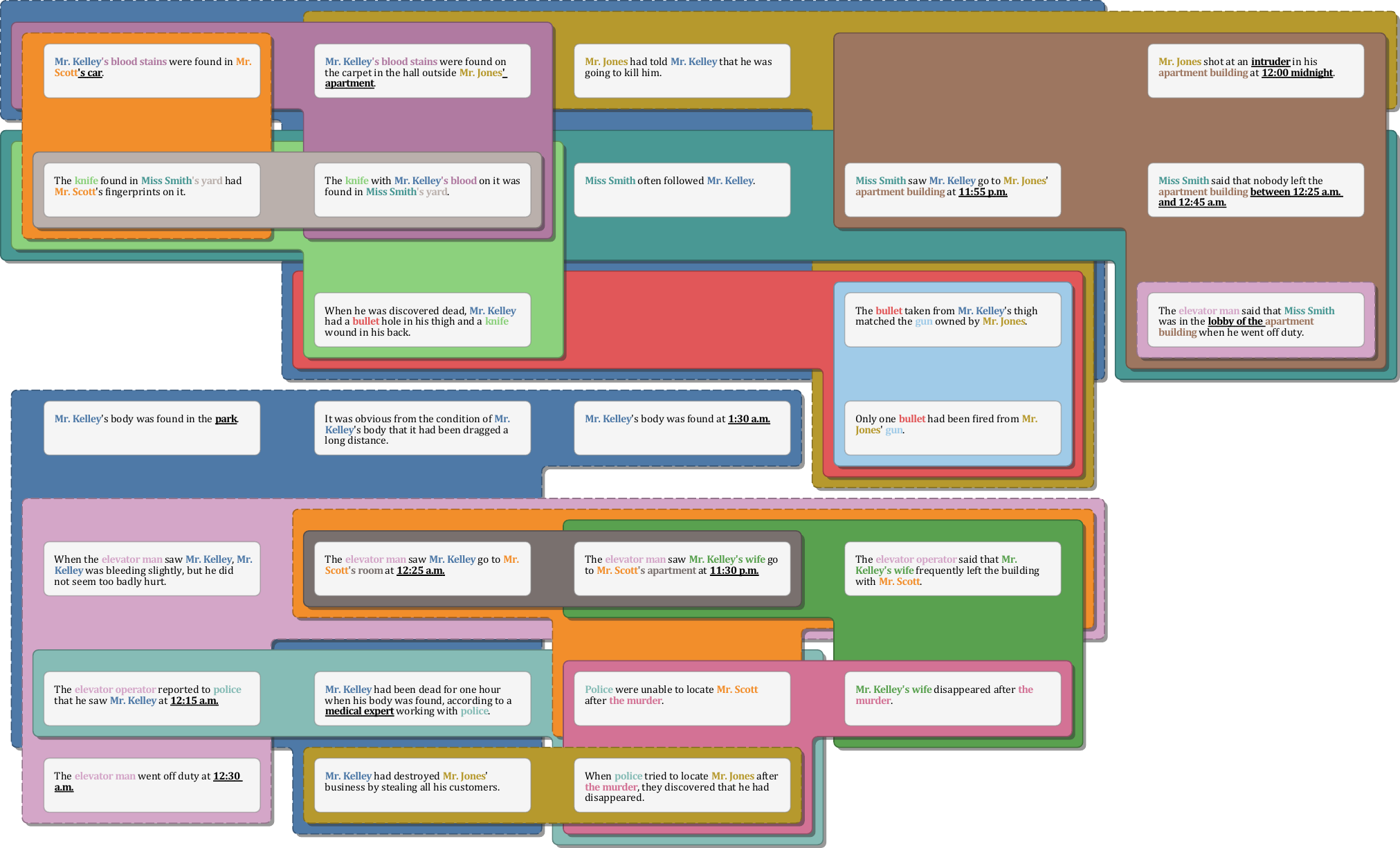}
        \caption{BlockSets with $\nabla$-shapes, split into two parts, the 4 repeated sets highlighted with dashed outline. Less compact than MosaicSets, more compact than RectEuler, more readable than both.}
        \label{fig:murder:BlockSets}
    \end{subfigure}
    \hfill
    \begin{subfigure}[t]{0.4\textwidth}
        \centering
        \raisebox{-4.5cm}{\includegraphics[width=\textwidth]{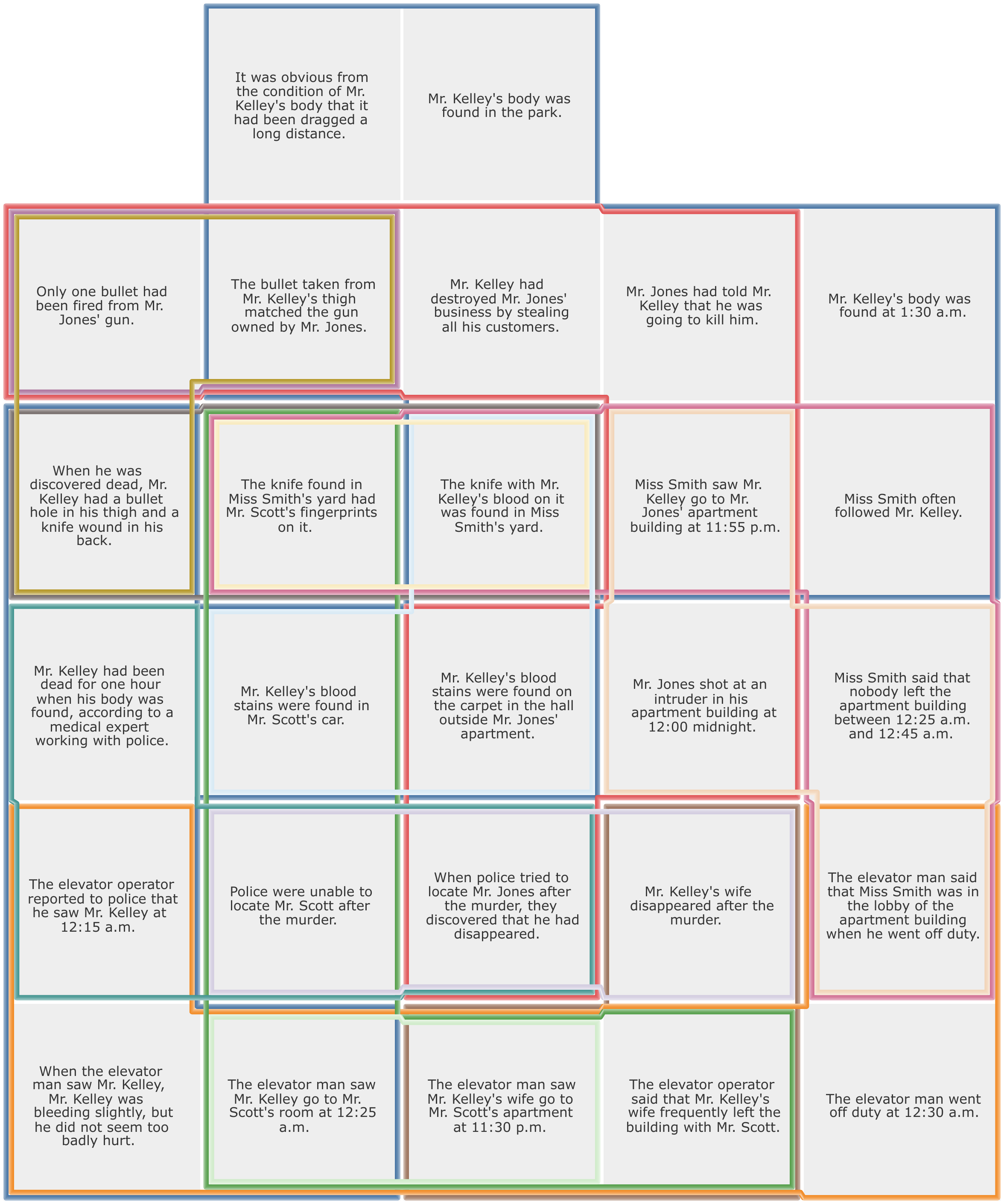}}
        \caption{MosaicSets~\cite{DBLP:journals/tvcg/RottmannWBGNH23}:  compact layout with arbitrary rectilinear shapes, crossing set boundaries, sets difficult to trace.}
        \label{fig:murder:MosaicSets}
    \end{subfigure}
    \hfill
    \caption{Statement-entity set system extracted from a murder mystery puzzle~\cite{stanford1969learning} using an LLM. RectEuler and MosaicSets computed with code provided by the authors. \autoref{fig:murder:BlockSeys:orthoconvex} in the supplementary material shows a contiguous orthoconvex BlockSets visualization of the same data set. Furthermore, \autoref{fig:murder:BlockSets-render:RectEuler} and \autoref{fig:murder:BlockSets-render:MosaicSets} show the RectEuler and the MosaicSets solutions rendered in the BlockSets style.
    }
    \label{fig:murder}
\end{figure*}

Though there are many methods to visualize sets systems, only few are able to accommodate large rectangular element shapes. These include RectEuler by Paetzhold \etal~\cite{paetzold2023recteuler} and the ``Untangled Euler Diagrams'' by Riche and Dwyer~\cite{DBLP:journals/tvcg/RicheD10}. Both approaches use rectangles to represent sets and are hence in principle able to incorporate larger blocks of data. However, neither imposes an alignment on the layout and the rectangular set shapes are not flexible enough to create compact visualizations. MosaicSets~\cite{DBLP:journals/tvcg/RottmannWBGNH23}, in contrast, places elements on a grid and uses arbitrarily complex rectilinear polygons as enclosing shapes. The visualization is structured and compact, but the polygon shapes are difficult to follow due to their high complexity (see \autoref{fig:murder}).

\mypar{Contributions.} In this paper we introduce {\bfseries BlockSets}: a novel set visualization technique which is specifically designed for sets with large elements. BlockSets creates a structured visualization by placing the elements on a grid and using orthoconvex rectilinear shapes as enclosing geometries. As such, our method covers the middle ground between the rectangular set representations of RectEuler and the arbitrary rectilinear shapes of MosaicSets. Orthoconvex shapes provide the necessary flexibility to find compact layouts while improving the readability. 
Set visualization techniques often use only outlines or a transparent fill to render the enclosing geometries. In contrast, BlockSets uses stacked opaque shapes. 
Doing so alleviates readability issues that are caused by crossing and coinciding set outlines and color-blending of transparent regions. Specifically, our algorithmic contributions are:
\begin{description}
\item[Layout Algorithm.] An integer linear program (ILP) that finds compact layouts of the set elements on a grid. The ILP supports general orthoconvex enclosing shapes, more restricted and hence visually simpler $\nabla$- and $\Gamma$-shapes, as well as rectangles.
    \item[Splitting Algorithm.] Not all set systems allow a compact contiguous representation on a grid and with orthoconvex enclosing shapes. Hence we present an algorithm that splits the set system in a suitable manner, solves the parts independently, and then compactly arranges the parts. Our algorithm generates fewer duplicate sets than the heuristic used by RectEuler~\cite{paetzold2023recteuler}. Our visual encoding highlights duplicate sets for the user.
    \item[Stacking Algorithm.] Finally, we describe an algorithm which computes a stacking order for our sets such that all shapes can be inferred despite the opaque rendering. While such a stacking does not have to exist for arbitrary set systems, our algorithm did find a solution for all real-world data sets that we tested.
\end{description}
\autoref{fig:pipeline-single-component} shows our algorithmic pipeline for a single component.

\mypar{Organization.} We review related work in the next section. Then, in \autoref{sec:design}, we discuss the specific design decisions and objectives underlying BlockSets. The following sections describe our algorithmic contributions in detail: the layout algorithm in \autoref{sec:layout-algorithm}, the splitting algorithm and the arranging of parts in \autoref{sec:splitting-algorithm}, and the stacking algorithm in \autoref{sec:stacking-algorithm}. We conclude the paper with a showcase of BlockSet visualizations and a Discussion.

\section{Related Work}

We give a brief overview of techniques related to BlockSets that either visualize sets or use grids.
For a comprehensive overview of set visualization, we refer the reader to the survey by Alsallakh \etal~\cite{DBLP:journals/cgf/AlsallakhMAHMR16}.
They classify set visualization into six categories.
BlockSets does not neatly fall within one such category, but is most related to Euler and Venn diagrams, and has some similarity to matrix-based techniques; hence, we discuss only set visualizations in these categories in this section.

\mypar{Euler and Venn diagrams.}
Euler and Venn diagrams use overlapping shapes to show the ways in which a collection of sets intersect.
Venn diagrams show all possible set intersections, while Euler diagrams aim to show only those set intersections that are present in the input data.
If an Euler diagram succeeds in staying true to the input data, then it is \emph{well-matched}.
An Euler diagram is \emph{well-formed} if its shapes intersect transversally, each set is represented by exactly one shape, and each set intersection forms exactly one connected zone in the drawing.
Though BlockSets has similarities to Euler diagrams, it is not our intention for it to be an Euler diagram.
In particular, BlockSets is not well-formed as we do not require set intersections to form a single connected zone and because sets may be represented by multiple shapes.
Furthermore, BlockSets is not well-matched; for example, in \autoref{fig:murder:BlockSets} there is a red-blue zone (where the bullet and Mr.\ Kelley entities overlap), but no statement belongs to exactly those two sets.
As the focus in BlockSets lies on the elements of the set system, we believe that this does not lead to confusion and we instead make use of the flexibility that non-well-matchedness provides.

Euler diagrams generally use curved shapes~\cite{chow2005towards, DBLP:journals/tvcg/KehlbeckGWD22, wybrow2021euler, perez2018nvenn, micallef2014eulerforce, rodgers2008general, stapleton2010inductively}, which hinders the integration of large visual representations of elements.
As mentioned in the introduction, RectEuler~\cite{paetzold2023recteuler}, MosaicSets~\cite{DBLP:journals/tvcg/RottmannWBGNH23} and the ``Untangled Euler Diagrams''~\cite{DBLP:journals/tvcg/RicheD10} are the Euler-like diagrams that are most suited to visualizing large elements (\autoref{fig:murder}).
There are also other Euler-diagram-like visualizations such as MetroSets~\cite{DBLP:journals/tvcg/JacobsenWKN21}.
Though the more regular layout of MetroSets helps with the integration of element representations, MetroSets visualizations do not use space efficiently which especially precludes their use for large elements.

There are some mathematical results on what class of shapes can represent which types of set systems.
For rectangles, it is known that the problem of recognizing intersection graphs of isothetic rectangles in the plane is NP-complete~\cite{kratochvil1994special}, which effectively implies that, unless P = NP, there are no efficient algorithms that can determine whether a set system can be represented by a well-matched Euler diagram that uses one rectangle to represent each set.
For convex polygons, it is known that one can represent arbitrarily large set systems where all sets pairwise intersect in a well-matched Euler diagram that uses one convex polygon per set but allows disconnected zones~\cite{grunbaum1975venn}.
This result extends to the orthoconvex polygons used by BlockSets.

\mypar{Hypergraph drawings.}
A set visualization can be viewed as a drawing of a hypergraph whose hyperedges are formed by the sets.
There is research on subdivision drawings~\cite{DBLP:conf/gd/KaufmannKS08, DBLP:conf/gd/BevernKKNS16} and planar supports of hypergraphs~\cite{DBLP:journals/jgaa/BuchinKMSV11}, which are related to Euler diagrams.
Drawings that use disjoint polygons to represent the sets have also been studied~\cite{DBLP:conf/walcom/DepianDKW24, DBLP:conf/gd/GoethemKKMSW17}.
Though related and theoretically interesting, this research does not provide algorithms to create a fully-functional set visualization.

\mypar{Matrix set visualizations.}
Matrix set visualizations are very compact and can visualize large-scale data.
They use a square grid to visualize sets, just like BlockSets.
However, they function quite differently, as the cells in the matrix encode binary information such as element membership~\cite{DBLP:journals/iwc/KimLS07, DBLP:journals/tvcg/SadanaMDS14, DBLP:journals/tvcg/WallingerDN23} or set-set intersection~\cite{DBLP:journals/tvcg/LexGSVP14}, rather than containing visual representations of the elements.

\mypar{Other grid visualizations.}
Outside set visualization, there are several grid visualizations that have similarities to BlockSets.
Yoghourdjian~\etal\ visualize networks on a grid and draw enclosing rectangles to form groups of nodes~\cite{yoghourdjian2015high}.
These groups allow for a simpler drawing: in their visualization, an edge between two groups implies that all nodes in the first group are connected to all nodes in the second.
Grids are also prevalent in the analysis of machine learning model output~\cite{DBLP:journals/tvcg/ZhouYCCSLYL24}, and in the visualization of small-multiples~\cite{DBLP:journals/tvcg/MeulemansSS21}.

\section{BlockSets}\label{sec:design}

In this section, we discuss the objectives and design of BlockSets. 
The input for a BlockSets visualization consists of elements  
and a collection of subsets of those elements.
Each element
has an associated visual representation such as a fragment of text, an image, or a chart.
We assume that the visual representation of each element fits in rectangles of similar size.

\mypar{Design objectives.}
Following Euler diagrams and the Gestalt principle of common region~\cite{gestalt}, we use enclosing geometries to represent sets.
Our two main design objectives are the following.
\begin{description}
\item[Compactness.] The visualization $(i)$ takes up little space in addition to the data elements; and $(ii)$ has small aspect ratio.
\item[Readability.]
The visualization is structured and
the user can efficiently determine $(i)$
for an element, all the sets to which it belongs; and $(ii)$
for a set, all elements that it contains.
\end{description}
Note that these objectives are in conflict: using less space for the sets generally impacts the readability of those sets negatively.

\mypar{Grid layout.}
We choose to arrange the elements on a rectangular grid.
Grid layouts are long accepted in typesetting and page layout as a basic design principle and their applicability and importance to graph and set data arrangement has also been explored by the visualisation community~\cite{yoghourdjian2015high, DBLP:journals/tvcg/RottmannWBGNH23}.
A grid layout results in a structured visualization as the rectangular representations of elements are aligned; it is also relatively compact.

\begin{figure}[t]
    \includegraphics[page=1, scale=0.75]{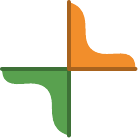}
    \hspace{2mm}
    \includegraphics[page=2, scale=0.75]{point-overlap.pdf}
    \hfill
    \includegraphics[page=3, scale=0.75]{point-overlap.pdf}
    \hspace{2mm}
    \includegraphics[page=4, scale=0.75]{point-overlap.pdf}
    \caption{Left: the types of point overlap we allow; colored area indicates interior. Right: we round to improve readability.}
    \label{fig:point-overlap}
\end{figure}

Naturally, BlockSets also places the enclosing shapes on the rectangular grid.
For readability, we require that no two edges of these enclosing shapes overlap in more than a point.
In particular, overlap in a single point is allowed only in two cases (\autoref{fig:point-overlap}).
To ensure non-overlapping shapes, we create gaps between consecutive rows and columns of the grid (\autoref{fig:pipeline-single-component}).
We round the corners of the enclosing shapes slightly to improve readability and to avoid the visual salience of sharp corners.
Edges of the enclosing shapes are spaced apart at a regular distance.

\begin{figure}[b]
\begin{subfigure}{0.475\linewidth}
    \includegraphics[width=\linewidth]{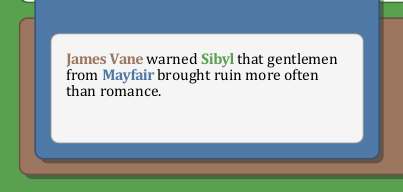}
    \caption{Colored text.}
\end{subfigure}
    \hfill
\begin{subfigure}{0.475\linewidth}
    \includegraphics[width=\linewidth]{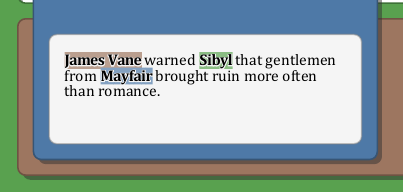}
    \caption{Colored background.}
\end{subfigure}
\caption{Two designs for highlighting entity names.}
    \label{fig:colored-sets}
\end{figure}

\mypar{Set representation.}
Set data may have a visual representation not only of the elements but also of the sets.
In statement-entity data, this is the name of the entity.
For compactness, we choose to not draw these names as labels, but instead highlight the entity names in the statement text with a single unique color and use the same color for the enclosing shape of the entity.
We considered two designs: colored text, and a colored background that acts as highlight (see \autoref{fig:colored-sets}).
Colored text requires dark colors, and, hence, works only if the number of non-singleton sets is relatively small. However, it generally appears less cluttered; all examples in this paper are rendered this way.
Note that singleton entities need neither color nor an enclosing shape; we simply use boldface and underline to highlight them in their statement.
Set systems other than statement-entity relations may benefit from a label; in \autoref{fig:EU:BlockSets} we show a mock-up of BlockSets with labels. We do not discuss label placement or its algorithmic challenges further in this paper. 

\mypar{Splitting.}
Some set systems do not admit a compact and readable visualization if each set is represented by a single shape.
Hence, in such cases, we allow the use of multiple shapes for a selection of sets.
To keep the problem computationally feasible, we follow RectEuler~\cite{paetzold2023recteuler} in splitting the visualization into separate parts that can be solved independently (see \autoref{fig:murder:BlockSets} for an example).
Each part has its own enclosing shapes, but some sets will be shared between the parts.
These sets can be matched to each other via their color.
In BlockSets, the shapes that represent shared sets receive a dashed boundary to make it clear to the user which sets are represented with multiple shapes.
Note that splitting the input allows any set system to be visualized, since in the limit each element has its own component.
An alternative to using multiple enclosing shapes is the duplication of elements as done by DupEd~\cite{DBLP:journals/tvcg/RicheD10}.
For large elements, duplicating their representation is expensive in terms of space; hence, we decided against this approach.

\mypar{Stacking.}
\begin{figure}
    \begin{subfigure}{\linewidth}
        \includegraphics[width=\linewidth]{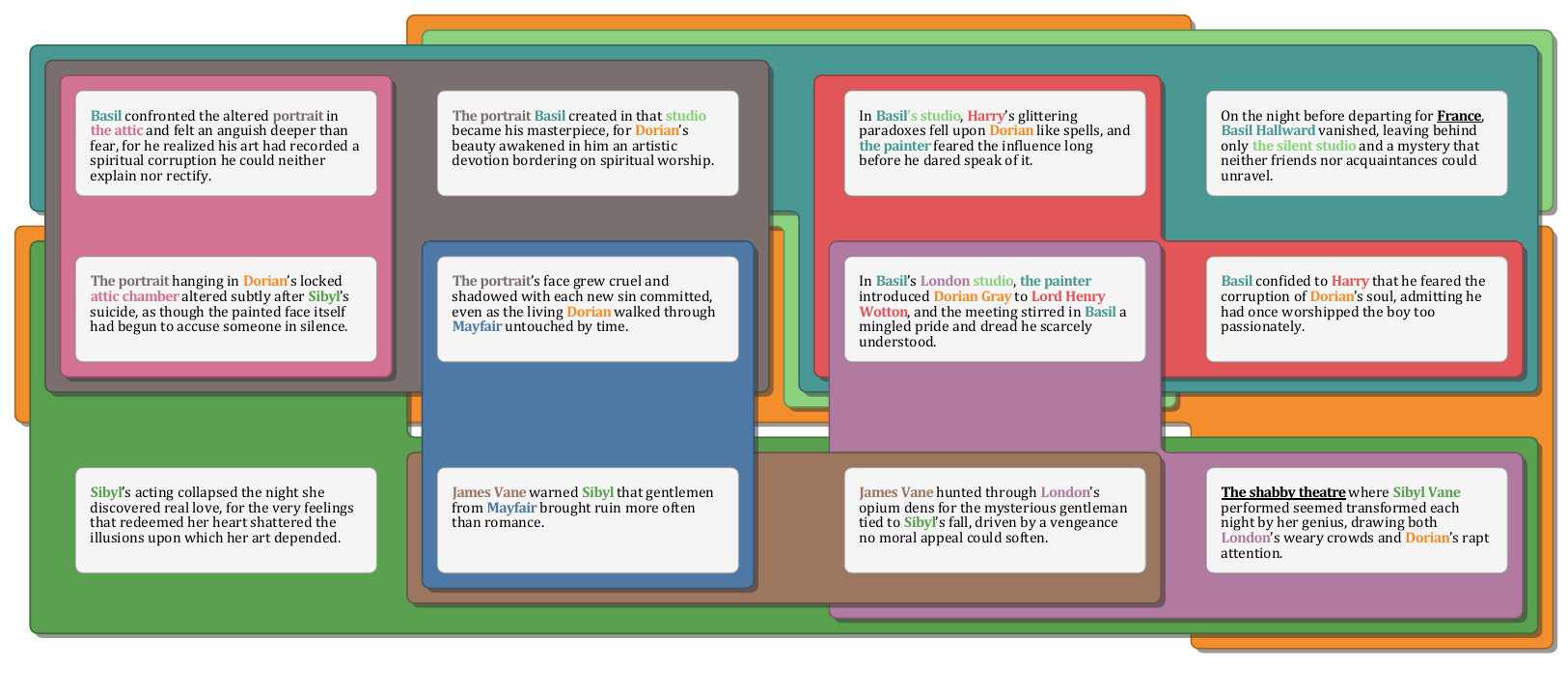}
        \caption{Opaque fill color.}
    \end{subfigure}

    \vspace{2mm}
    
    \begin{subfigure}{\linewidth}
        \includegraphics[width=\linewidth]{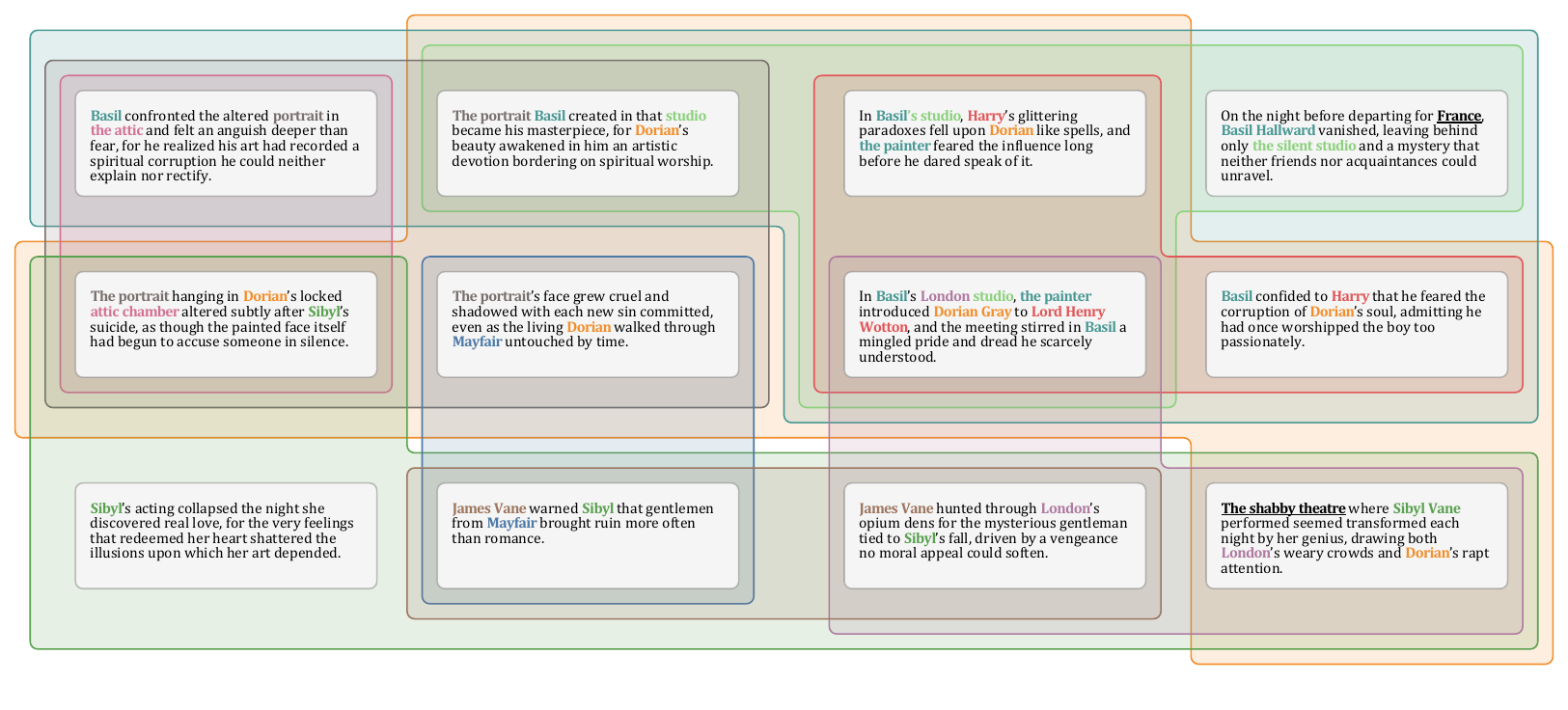}
        \caption{Transparent fill color.}
    \end{subfigure}
    \caption{Two BlockSets renderings of ``Wilde-style'' narrative statements inspired by ``The Picture of Dorian Gray'' and generated with GPT 5.1.}
    \label{fig:wilde-opaque-transparent}
\end{figure}
Most set visualizations use only outlines or a transparent fill to render the enclosing geometries~\cite{BubbleSets, paetzold2023recteuler,DBLP:journals/tvcg/KehlbeckGWD22, DBLP:journals/tvcg/RottmannWBGNH23}.
Outlines only can make it difficult to determine for an element all sets to which it belongs (see also \autoref{fig:murder:MosaicSets}).
Transparency alleviates these issues somewhat, but causes color blending in regions with multiple overlapping sets, which can leave the visualization a unappealing and ineffective muddy brown in many places (see \autoref{fig:wilde-opaque-transparent} bottom). We hence choose an opaque fill color (see \autoref{fig:wilde-opaque-transparent} top).

Though opaqueness more easily facilitates color matching, the shapes representing sets can now cover each other, and the stacking of shapes becomes important.
This brings the question: can we find a stacking that ensures all shapes can be inferred?
In \autoref{sec:stacking-algorithm}, we describe an algorithm that finds such a stacking if one exists.
Though one can construct examples for which a stacking does not exist, we have not encountered any such case in the real-world data that we tested.
If such a situation should arise, then one can add visual aids to ensure that the shapes can be inferred.

Note that also the Untangled Euler diagrams of Riche and Dwyer~\cite{DBLP:journals/tvcg/RicheD10} use opaque colors to render their sets. However, their design in essence stacks only shapes that are nested and hence they do not need to resolve issues revolving around covered set boundaries. We are not aware of previous set visualization work that stacks shapes in the way that we propose for BlockSets.
Cabello \etal~\cite{proportional-symbol-maps} described algorithms to create clear drawings of proportional symbol maps that use opaque symbols such as disks or squares.
They aim to maximize the visible boundary length of each symbol, and consider two types of drawings: stacking drawings, and physically realizable drawings where shapes are allowed to ``bend'' in the third dimension.
We consider only stacking drawings as physically realizable drawings can be confusing to the user and stacking drawings already allow all shapes to be inferred.

\mypar{Set color.}
BlockSets relies on color to (1) distinguish sets from each other; $(2)$ visually group repeated sets in case the input is split; and $(3)$ match entity names to the corresponding enclosing shape for statement-entity data.
We use Tableau20 as color palette and simulated annealing as a heuristic to find an assignment of color to sets such that sets that are spatially close receive colors that are as distinct as possible according to their CIEDE2000 color difference. 
Here we consider sets that overlap to be considerably closer than shapes that do not overlap. Furthermore, overlapping sets next to each other in the stacking order are closer than those further apart.

\begin{figure}[t]
    \centering
    \includegraphics{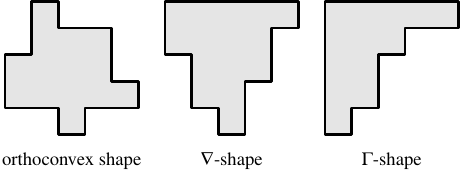}
    \caption{Enclosing shapes used by BlockSets.}
    \label{fig:enclosing-shapes}
\end{figure}

\begin{figure*}[b]
    \centering
    \includegraphics[page=1]{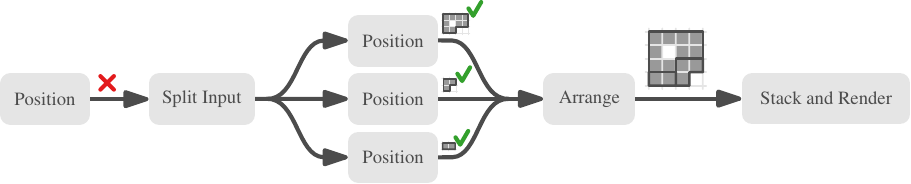}
    \caption{If no compact assignment of elements to grid cells exists, BlockSets splits the input into parts (\autoref{sec:splitting-algorithm}). The elements of each part are first positioned separately (\autoref{sec:layout-algorithm}), then arranged compactly in a single grid (\autoref{subsec:arrange}), then the sets are stacked and rendered.}
    \label{fig:split-pipeline}
\end{figure*}

\mypar{Type of enclosing shape.}
The choice of enclosing shape has a large effect on the compactness and readability of the visualization.
Rectangles as used by \cite{DBLP:journals/tvcg/RicheD10, paetzold2023recteuler} are the simplest shape to read, but give no flexibility.
In contrast, arbitrary rectilinear polygons as used by MosaicSets \cite{DBLP:journals/tvcg/RottmannWBGNH23} are flexible, but hard to read.
As an illustration, consider the enclosing shape for Mr. Kelley in \autoref{fig:murder:MosaicSets}: The second column switches between statements that are included in the set, and not included in the set four times. These types of ``gaps'' make it difficult to see the entirety of the set at a glance.
The shapes that avoid such switches are \emph{orthoconvex} shapes (short for orthogonally-convex shapes), which are those shapes for which the intersection of any vertical or horizontal line with the shape is either empty or consists of a single component.

Rectangles generally do not lead to compact set visualizations, but their straight edges make it easy for the reader to infer the global structure of the shape from local features. 
For BlockSets we want to use the added flexibility of orthoconvex shapes, but still retain parts of the helpful structure of rectangles.
We call a shape \emph{top-aligned}, if for any point $(x, y)$ in the shape, a line segment from $(x, y)$ to $(x, y_\mathit{max})$ is fully contained in the shape; here $y_\mathit{max}$ denotes the largest $y$-coordinate of the shape.
We similarly define \emph{left-aligned}.
We consider the following types of shapes, from least to most restrictive (\autoref{fig:enclosing-shapes}):
\begin{itemize}
    \item arbitrary rectilinear shapes;
    \item orthoconvex shapes;
    \item \emph{$\nabla$-shapes}: orthoconvex shapes that are top-aligned;
    \item \emph{$\Gamma$-shapes}: orthoconvex shapes that are top- and left-aligned;
    \item rectangles.
\end{itemize}
Top- and left-alignment were chosen to match the English left--right top--bottom reading order.
We find that orthoconvex and $\nabla$-shapes provide a good trade-off between compactness and readability.

\mypar{Computational pipeline.}
\autoref{fig:pipeline-single-component} shows our pipeline to compute a single component of BlockSets.
For complex set systems or restrictive enclosing shapes no compact assignment of elements to grid cells might exist. 
In such a case, BlockSets will split the input into parts, position them separately and arrange their solutions (\autoref{fig:split-pipeline}).
The following sections describe the algorithmic details of the steps in the pipeline.

\section{Layout Algorithm}\label{sec:layout-algorithm}
This section describes the most computationally intensive task: finding a compact assignment of elements to grid cells.
We use integer linear programming (ILP) to solve this problem optimally under a set of \emph{constraints} and for a certain \emph{objective function}.
For brevity, we provide a high-level overview of the approach and omit the formulas for the constraints and objective function; for details, we refer to the reader to the implementation\footnote{\href{https://github.com/tue-alga/BlockSets}{github.com/tue-alga/BlockSets}}.
We describe only the ILP for orthoconvex shapes, $\nabla$-shapes, and $\Gamma$-shapes.
For modeling rectangles and arbitrary polygons, we refer the reader to RectEuler~\cite{paetzold2023recteuler} and MosaicSets~\cite{DBLP:journals/tvcg/RottmannWBGNH23} respectively.

\begin{figure}[t]
    \centering
    \includegraphics[]{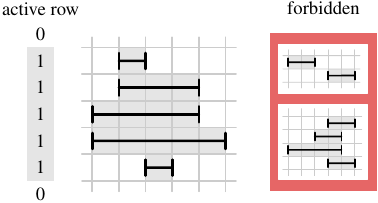}
    \caption{We model an orthoconvex polygon by intervals in a contiguous set of active rows.}
    \label{fig:ILP-constraints}
\end{figure}

\mypar{Element and set representation.}
The ILP works on a square grid of fixed size inside which all elements are positioned.
The size of the grid can be set to, for example, $\lceil\sqrt{n}\rceil + 1 \times \lceil\sqrt{n}\rceil + 1$, where $n$ is the total number of elements.
Boolean variables determine for each pair of an element and grid cell whether the element is placed in that grid cell.
Appropriate constraints ensure that an element is assigned to exactly one grid cell.
Each set is modeled by a set of intervals (\autoref{fig:ILP-constraints})
and uses at most one interval per row.
All rows with an interval are called \emph{active}.
For a set to be represented by an orthoconvex polygon:
\begin{enumerate}
    \item its active rows should be contiguous;
    \item its intervals of consecutive rows should overlap; and
    \item its intervals should form a vertically convex polygon.
\end{enumerate}
The last constraint means that for a given interval, it is forbidden that an interval above it and an interval below it both end later or both start earlier.
See \autoref{fig:ILP-constraints} for an illustration.

\mypar{$\nabla$- and $\Gamma$-shapes.}
To model $\nabla$-shapes, we add a constraint that each interval is a subset of the one above it.
For $\Gamma$-shapes, one variable determines for a given set the start of all of its intervals.

\mypar{Element membership.}
ILP constraints ensure that an element is placed within one of the intervals representing a set if and only if the element is part of that set.
Lastly, we do not allow two shapes to overlap if their sets do not share an element; hence, we require that the intervals of such sets that share a row are disjoint.

\mypar{Objective function.}
We minimize a linear combination of the following three terms (with equal weight):
\begin{description}
\item[Polygon area.]
For readability, we avoid empty grid cells within sets.
We model this by minimizing the sum of the polygon areas. 
\item[Total bounding box dimensions.]
For compactness, we want a visualization that takes up little space and has low aspect ratio.
Hence, we minimize the sum of the dimensions of the filled grid.
\item[Polygon complexity.]
Lastly, for readability we minimize the number of vertices of each polygon.
\end{description}

\section{Splitting Algorithm}\label{sec:splitting-algorithm}
If no compact layout is found by the ILP (\autoref{sec:layout-algorithm}), BlockSets splits the input into parts. Also RectEuler splits the input as needed; below we explain how our splitting approach differs from theirs. 
Ideally, we split the input such that each part can be visualized in the most compact and readable way.
However, there are many possible splits and it is computationally infeasible to run the layout algorithm on all these splits.
Hence, we require a heuristic approach.
RectEuler~\cite{paetzold2023recteuler} recursively clusters elements into two groups based on the sets to which they belong (via the Jaccard metric).
We found that this approach does not always lead to good splits, see Figures~\ref{fig:bank-robbery} and \ref{fig:bank-robbery:RectEuler:no-split} in the supplementary material.
Depending on the set time limit, RectEuler either does not split, which results in a very sparse visualization (\autoref{fig:bank-robbery:RectEuler:no-split}), or it does split and duplicates many sets.
This is due to the fact that RectEuler focuses on the elements and not the sets when computing splits.
However, duplicated sets negatively impact readability. We aim to minimize their number by taking a set-centric splitting approach.

\mypar{Deciding when to split.}
BlockSets uses a user-specified time limit for the ILPs, as in particular the ILP for the layout algorithm may take a long time to complete while it may find the optimal solution, or one close to it, quite quickly.
We split the input if the layout algorithm is not able to find a layout within this limit.
We also split the input if the layout returned by the algorithm is not compact.
We measure compactness by counting the number of empty grid cells within the \emph{orthoconvex hull} of the set of filled grid cells, which is the smallest orthoconvex shape that contains the set of filled grid cells.
We split if this number exceeds $\frac{1}{5}n$, where $n$ is the number of elements in the layout.
When using rectangles as enclosing shapes, there will generally be many empty grid cells; to not cause too many splits one can use a larger threshold or remove this splitting condition entirely.

\mypar{Quality of a split.}
Aside from duplicating few sets, we also aim for our split to make each part after the split as simple as possible.
To model this, we look for splits that spread the sets roughly evenly over the parts.
Hence, a split should result in few components that:
\begin{itemize}
    \item have as few duplicate sets as possible;
    \item are balanced (in terms of size); and
    \item require no further splitting.
\end{itemize}

\begin{figure}[b]
    \centering
    \includegraphics[page=3]{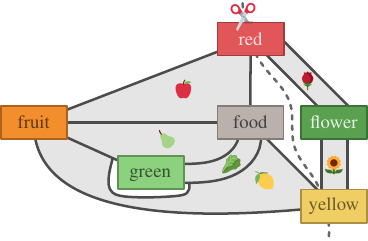}
    \caption{Zykov drawing of the dual hypergraph of the input of \autoref{fig:pipeline-single-component} (each hyperedge is a face in the drawing; all vertices incident to the face are part of the hyperedge). Vertex set $\{\text{red}, \text{yellow}\}$ forms a vertex separator.}
    \label{fig:intersection-graph}
\end{figure}

\begin{figure}[t]
    \centering
    \includegraphics[page=4]{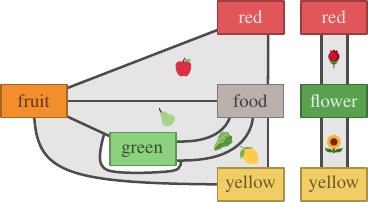}
    \caption{Dual hypergraph after split.}
    \label{fig:split-intersection-graph}
\end{figure}

\mypar{Dual hypergraph.}
We find splits via the dual hypergraph $G$ of the set system, which has a vertex for each set and a hyperedge for each element which connects all sets the element is part of (\autoref{fig:intersection-graph}).
We call a split \emph{proper} if after splitting each part contains at least one set that is not duplicated.
A proper split disconnects~$G$ by removing a set of vertices.
That is, a proper split is based on a \emph{vertex separator} in~$G$.
Given a vertex separator $S$, we split as follows.
Vertex set $S$ separates $G$ into connected components $C_1, \dots, C_k$.
To create $G_1, \dots, G_k$ that form the BlockSets input parts after the split, we ``cut'' each vertex $v \in S$: we place a copy of the vertex in each of the components it connects to, and distribute its incident hyperedges accordingly over the components (\autoref{fig:split-intersection-graph}).
The only hyperedges that are not distributed are those that connect only vertices in~$S$.
The elements that correspond to these hyperedges can be placed freely in any component.
We place these iteratively, in decreasing order of hyperedge size, by putting them in the smallest component with the largest number of vertices from its hyperedge, and adding missing vertices to the component if needed.

The separator should remove as few vertices as possible such that the resulting components in the remaining graph are of similar size: the maximum size of a component is $\alpha n$ where $n$ is the number of vertices in the hypergraph.
Such separators are called minimum $\alpha$-vertex separators, and finding or approximating one in a regular graph is NP-hard~\cite{vertex-separator}.
Furthermore, a minimum vertex separator may not form a good split.
Consider a star graph.
The center vertex forms a vertex separator of size one.
The removal of the center vertex results in many components.
A better split would merge some of the components into larger pieces, such that their shared nodes need to be represented only once.
Therefore, to construct separators, we use a brute-force algorithm that iteratively merges the smallest two components until all components fall within our size limit.

\mypar{Brute-force algorithm.}
We iterate over all vertex separators of size at most 5 such that the size of each component after merging lies within $[\frac{1}{3}n, \frac{2}{3}n]$, where $n$ denotes the total number of vertices in the hypergraph.
The upper limit on separator size serves to restrict the enumeration to small separators.
We choose the vertex separator that minimizes the weighted sum of the following:
\begin{itemize}
    \item the number of components;
    \item the number of deleted vertices;
    \item the number of copies of deleted vertices across all components;
    \item 10 times the number of components whose size is above the maximum threshold (minimizing the possibility of a component requiring another split to be visualized); and
    \item the difference in size between the biggest and smallest component (ensuring balanced components).
\end{itemize}
If no vertex separator is found by the brute-force algorithm then we duplicate all sets.
In this case, we create two empty components; all hyperedges have vertices only in the separator $S$ and they are distributed among the components in decreasing size.

\subsection{Arranging Components}\label{subsec:arrange}
If a BlockSets visualization is split, we compute layouts for each component.
These then need to be arranged in a compact way.
This optimization problem can, just like the layout problem, be solved using an ILP.
The two main constraints for a valid solution are:
\begin{enumerate}
    \item each component is placed exactly once; and
    \item there is no overlap between components.
\end{enumerate}
We allow mirroring and rotation of the components if the shape class we use is closed under that operation.
That is, we do not rotate or mirror vertically when using $\nabla$- and $\Gamma$-shapes and we do not mirror horizontally when using $\Gamma$-shapes.
We also do not mirror rectangles nor rotate them 180 degrees.
We only use a transformed component if its shape is different from other versions of the component.
We define a boolean variable for each pair of component and position it can be placed.
For each component, exactly one of these variables should be set to true, ensuring the first constraint is satisfied.
For the second constraint, we count for each cell of the grid how many occupied cells are placed at this position, and constrain it to be at most one.

\begin{figure}
    \begin{subfigure}{0.24\linewidth}
        \centering
        \includegraphics[page=3]{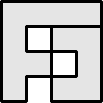}
        \caption{}
    \end{subfigure}
    \hfill
    \begin{subfigure}{0.24\linewidth}
        \centering
        \includegraphics[page=2]{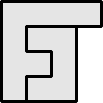}
        \caption{}
    \end{subfigure}
    \hfill
    \begin{subfigure}{0.24\linewidth}
        \centering
        \includegraphics[page=4]{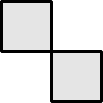}
        \caption{}
    \end{subfigure}
    \hfill
    \begin{subfigure}{0.24\linewidth}
        \centering
        \includegraphics[page=5]{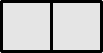}
        \caption{}
    \end{subfigure}
    \caption{Arrangements (a) and (b) have equal bounding box dimensions, but (b) has smaller orthoconvex hull area. Arrangements (c) and (d) have equal orthoconvex hull area, but (d) has smaller bounding box dimensions.}
    \label{fig:arranging:objective-function}
\end{figure}

\mypar{Objective function.}
We minimize a linear combination of the following terms.
\begin{description}
\item[Total bounding box dimensions.]
We minimize the sum of the width and height of the bounding box dimensions.

\item[Dimension difference.]
To avoid excessively stretched layouts, we minimize the absolute difference between the width and height.

\item[Orthoconvex hull area.]
Among solutions that have roughly the same bounding box, we prefer the one where the components are arranged in a more compact manner (\autoref{fig:arranging:objective-function}).
To achieve this preference we add a term for the orthoconvex hull area.

\end{description}

\section{Stacking Algorithm}\label{sec:stacking-algorithm}
\begin{figure}[tb]
    \centering
    \includegraphics{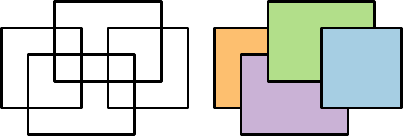}
    \caption{In any stacking at least one of the rectangles is covered.}
    \label{fig:stackingCounterexample}
\end{figure}

BlockSets renders enclosing geometries opaquely and hence they can obscure each other. We compute 
a stacking order such that each shape is as readable as possible.
It is generally not necessary for the entirety of a shape to be visible, as 
long as it can be inferred from the visible portions.
For rectilinear polygons we define this kind of visibility as follows: a polygon (set) is visible if and only if all of its edges are visible.
An edge is visible if some part of it is visible.
If a polygon or edge is not visible then it is considered covered. 
It may not be possible to stack polygons such that they are all visible (see the  synthetic example in \autoref{fig:stackingCounterexample}), but our algorithm found a suitable stacking for all real-world examples we tested.

\subsection{Algorithm} \label{stackingAlgorithm}
Our stacking algorithm takes as input the shapes (polygonal boundaries of occupied cells) as constructed by the layout algorithm
and then proceeds to iteratively stack these polygons, bottom-up.
Intuitively, we always place the polygon next which is least covered by all remaining polygons still to be placed.
To this end, the algorithm maintains a set $S$ of polygons that have not been placed yet, which is initially the entire set of polygons.
Then it repeatedly performs the following steps. 
\begin{enumerate}
    \item Check for each polygon $P \in S$ for all its edges whether they are covered by the set of polygons $S \setminus \{P\}$.
    \item Place the polygon that has:
    \begin{enumerate}
        \item most visible edges; or, as a tie-breaker:
        \item most edges; or, as a tie-breaker:
        \item largest area.
    \end{enumerate}
\end{enumerate}

\mypar{Edge visibility.}
To perform the visibility check, we first compute the arrangement (intersection) of all polygons. Our final rendering step separates edges that are coinciding to render the sets as properly stacked upon each other. For this purpose we are reserving space in the grid early on in our pipeline (recall Step 1 in \autoref{fig:pipeline-single-component}: the numbers indicate the edges of shape polygons coinciding at that position). Hence, for our purpose, two edges that fully overlap are both considered visible; an edge is covered only if it is covered by the proper interior of another polygon. Similar reasoning applies to vertices of the arrangement: they are covered only if they lie in the proper interior of another polygon or lie on the boundary of another edge in a configuration other than the two cases shown in \autoref{fig:point-overlap}.

\begin{figure*}[b]
    \begin{subfigure}{0.33\textwidth}
        \includegraphics[width=\linewidth]{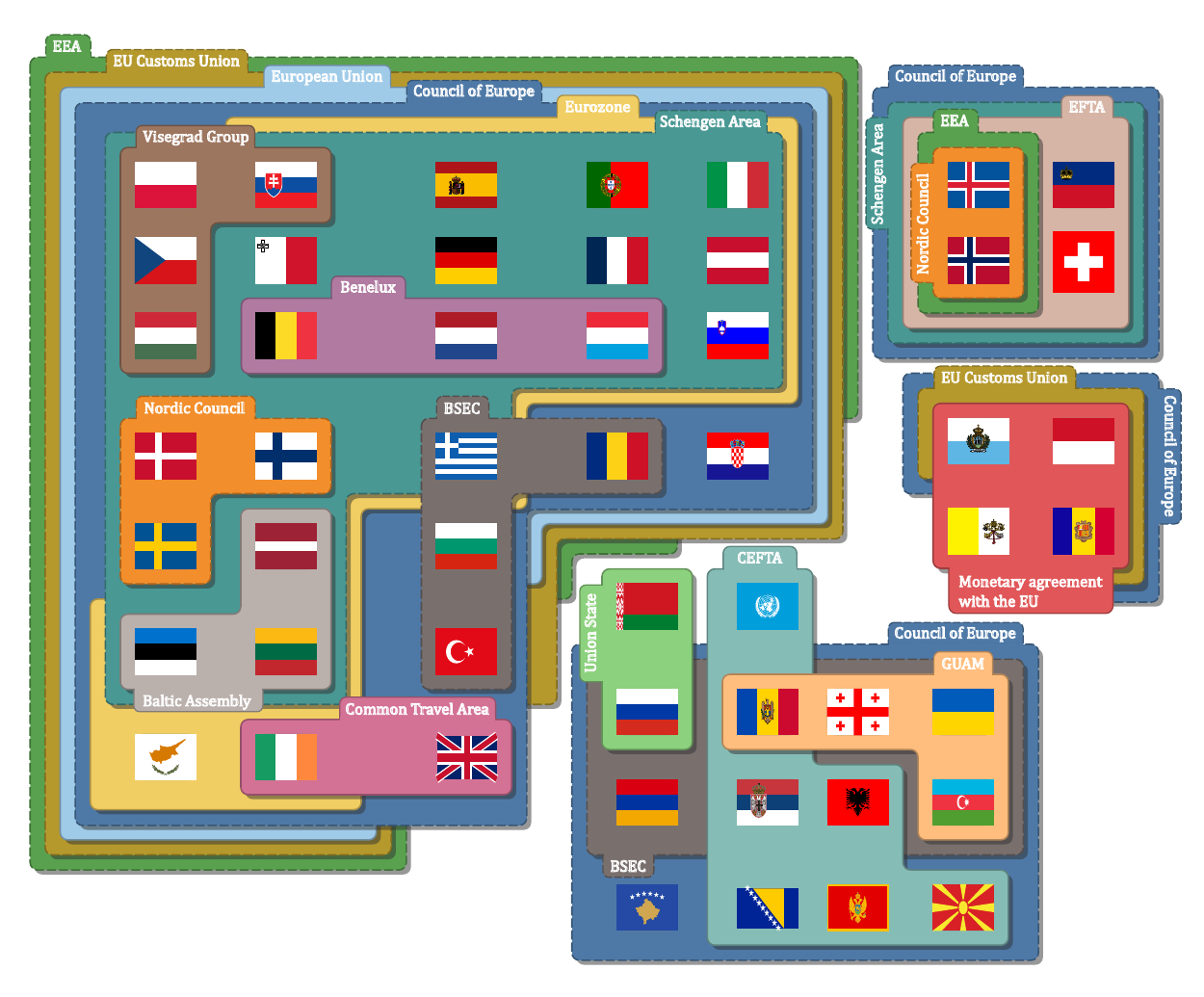}
        \caption{BlockSets with $\nabla$-shapes}
        \label{fig:EU:BlockSets}
    \end{subfigure}
    \hfill
    \begin{subfigure}{0.33\textwidth}
        \includegraphics[width=\linewidth]{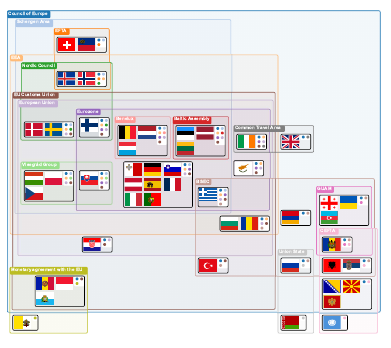}
        \caption{RectEuler~\cite{paetzold2023recteuler}}
    \end{subfigure}
    \hfill
    \begin{subfigure}{0.33\textwidth}
        \includegraphics[width=\linewidth]{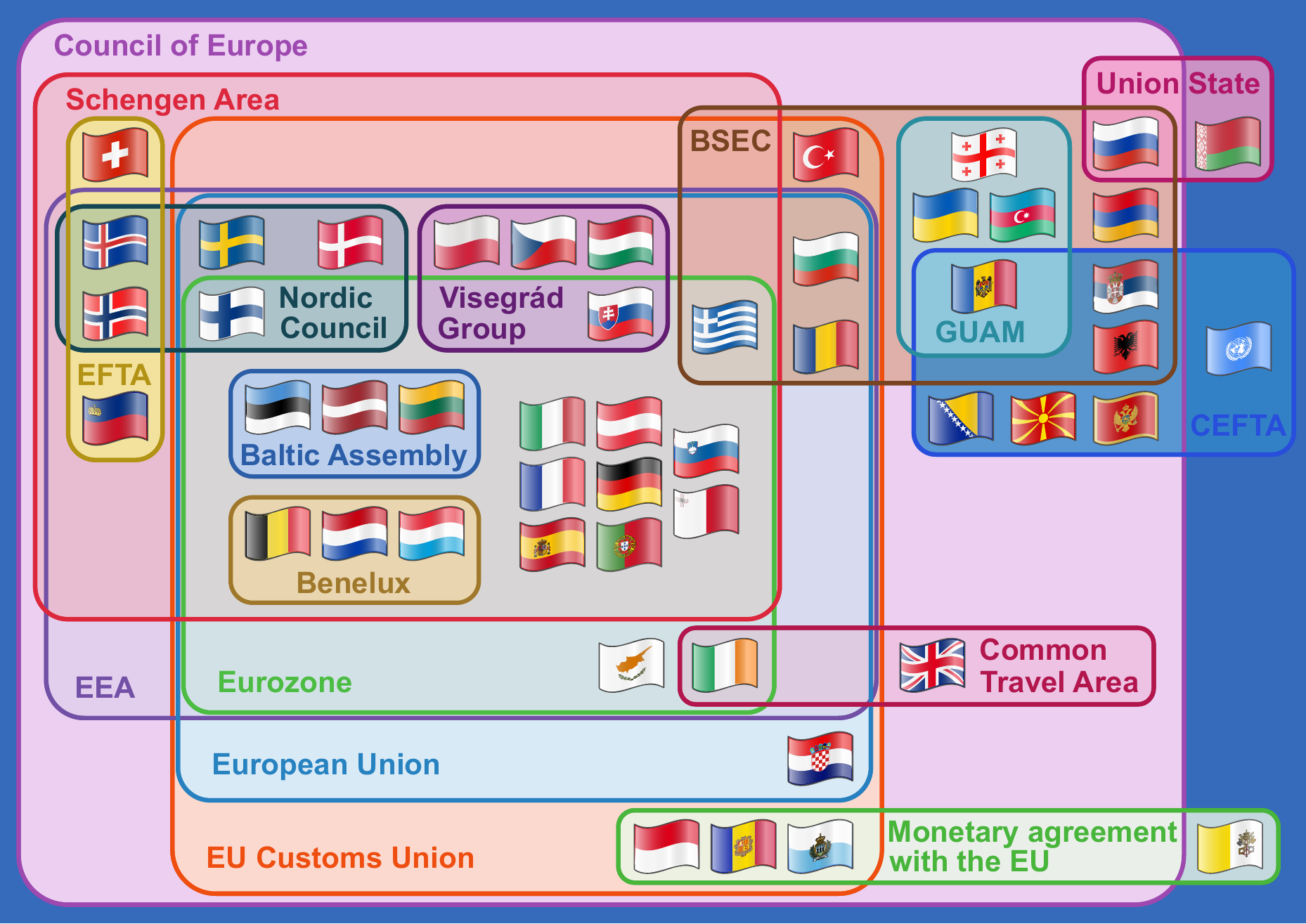}
        \caption{Handmade~\cite{EU}}
    \end{subfigure}
    \caption{A comparison of set visualizations of the supranational European bodies.}
    \label{fig:EU}
\end{figure*}

\section{Showcase}\label{sec:showcase}
In this section, we showcase BlockSets on a variety of datasets and using the range of different enclosing shapes.
We generated all figures using our prototype implementation of BlockSets, which is publicly available at \href{https://github.com/tue-alga/BlockSets}{github.com/tue-alga/BlockSets}.

\mypar{Statement-entity data.}
\autoref{fig:teaser}, \autoref{fig:murder}, \autoref{fig:wilde-opaque-transparent} show BlockSets on different types of statement-entity data: a collection of tales, a murder mystery, and a single story respectively.
The Robin Hood tales and the murder mystery have no given linear order, while the story of Dorian Gray has an underlying narrative.
Therefore, the use of BlockSets for these visualization will differ.
Though the tales and murder mystery statements have no given order, they would still traditionally be presented in a linear fashion.
BlockSets instead provides a two-dimensional reading experience.
One can start by reading the statements sequentially in the grid; the layout algorithm of BlockSets ensures that statements that are close to each other in the grid are similar to each other in terms of the entities they mention.
If a question about an entity arises, the reader can match the color of the entity to the enclosing shape(s) in the visualization and effectively find other statements that mention this entity.

Furthermore, the enclosing shapes allow the reader to effectively determine the prevalence of statements and entities.
In \autoref{fig:teaser} for example, the shape representing Robin Hood is large and covers all but three statements, showing his prevalence in these tales.
Turning to the visualization of Dorian Gray, the regular grid layout, together with the margins between enclosing shapes, makes one central statement pop out to the reader in importance: the meeting between Basil, Dorian Gray, and Lord Henry Wotton in Basil's London studio.
This also shows the promise of BlockSets for the analysis of stories: it provides a deconstructed view that brings to light the prevalence and interactions of entities.

\begin{figure}
    \centering
    \includegraphics[width=\linewidth]{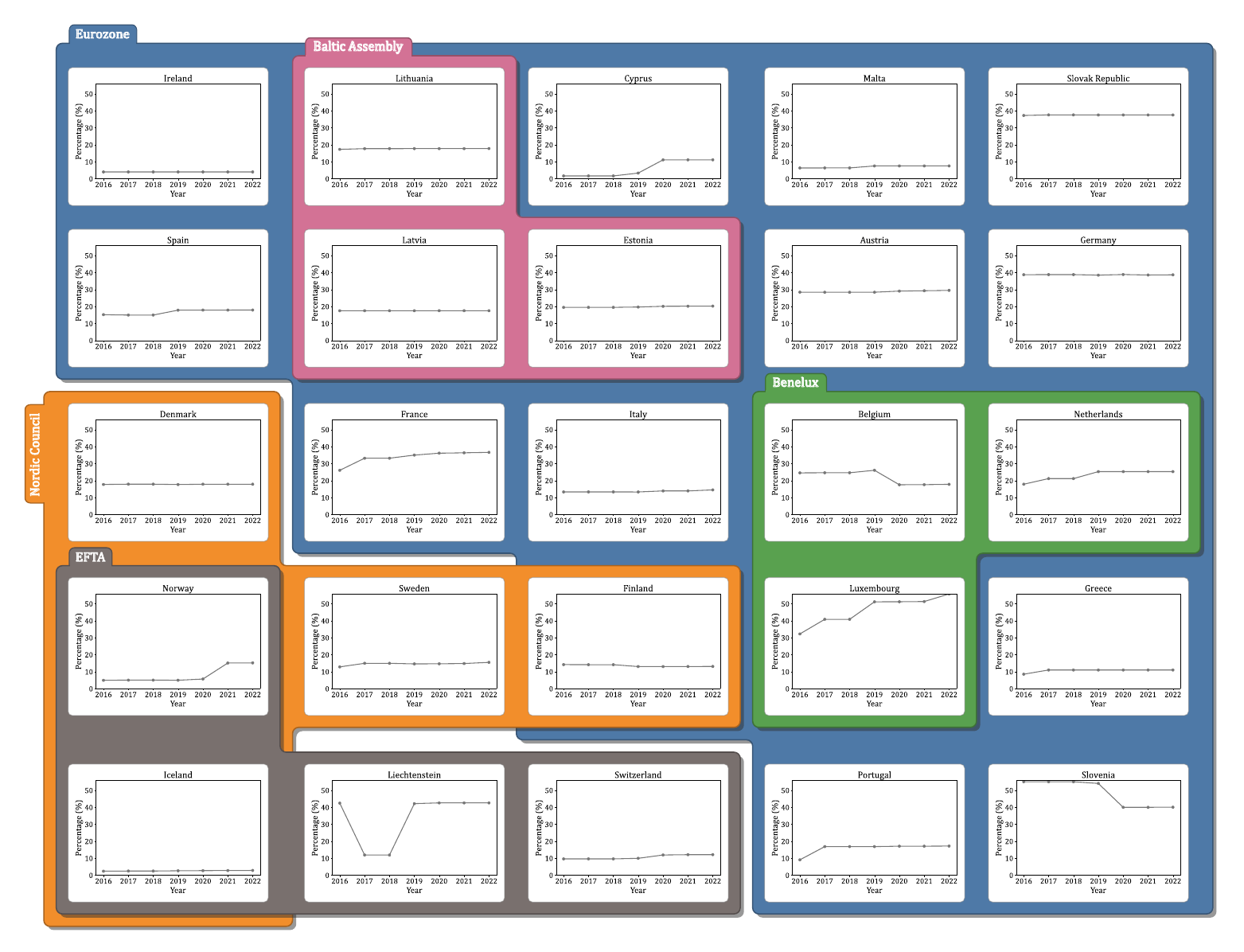}
    \caption{Using BlockSets to visualize terrestrial and marine protected areas as percentage of total area, 2016–2022. Source: World Bank, Environment, Social and Governance (ESG) Data, Indicator WB\_ESG\_ER\_PTD\_TOTL\_ZS.}
    \label{fig:small-multiple}
\end{figure}

\mypar{Small multiples and images.}
\autoref{fig:EU} and \autoref{fig:small-multiple} use BlockSets to visualize images and charts that have an underlying set system.
Here, we attached labels manually to the enclosing shapes to show the set names.
\autoref{fig:EU} shows a BlockSets visualization of the supranational European bodies in comparison to RectEuler and a manual Euler diagram.
As the elements here are small, we do not require alignment of elements in different components, so we arranged them manually rather than using the arranging algorithm of \autoref{subsec:arrange}.
Using the flexibility of $\nabla$-shapes and by splitting the input into parts, BlockSets was able to find a much more compact layout for the elements than RectEuler.

\begin{figure}
        \includegraphics[width=\linewidth]{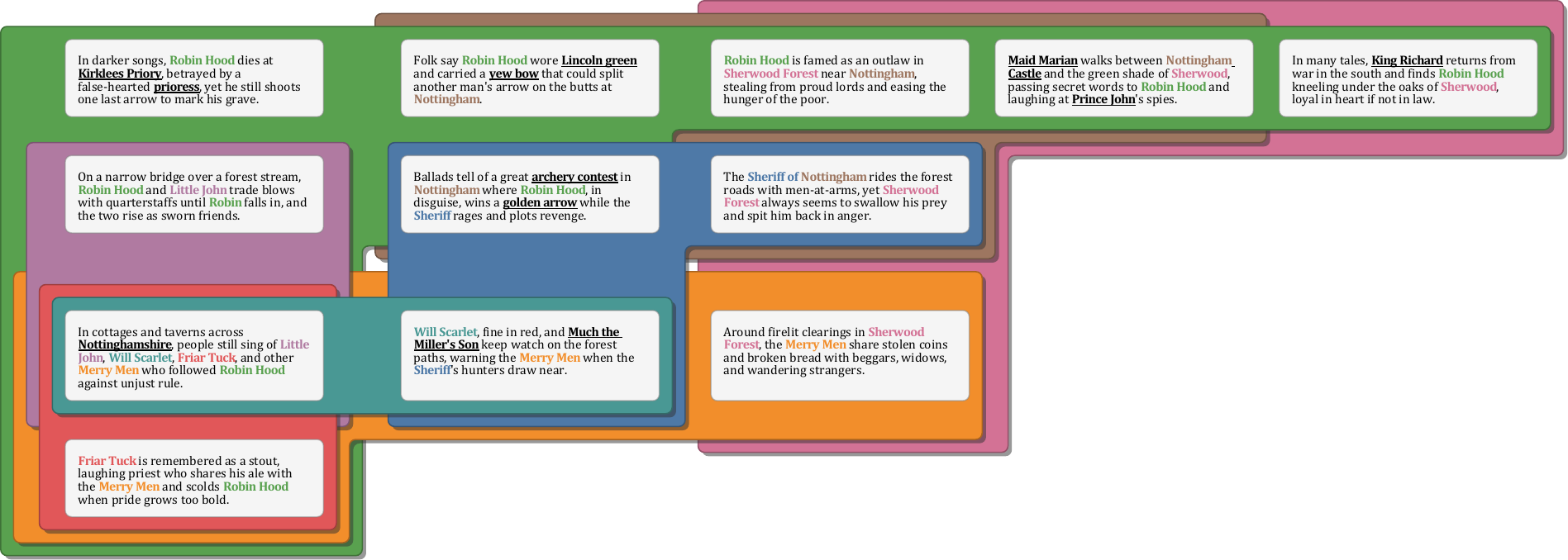}
        \caption{BlockSets with $\Gamma$-shapes for the Robin Hood tales.}
        \label{fig:robin-hood:gamma}
\end{figure}

\begin{figure}
        \includegraphics[width=\linewidth]{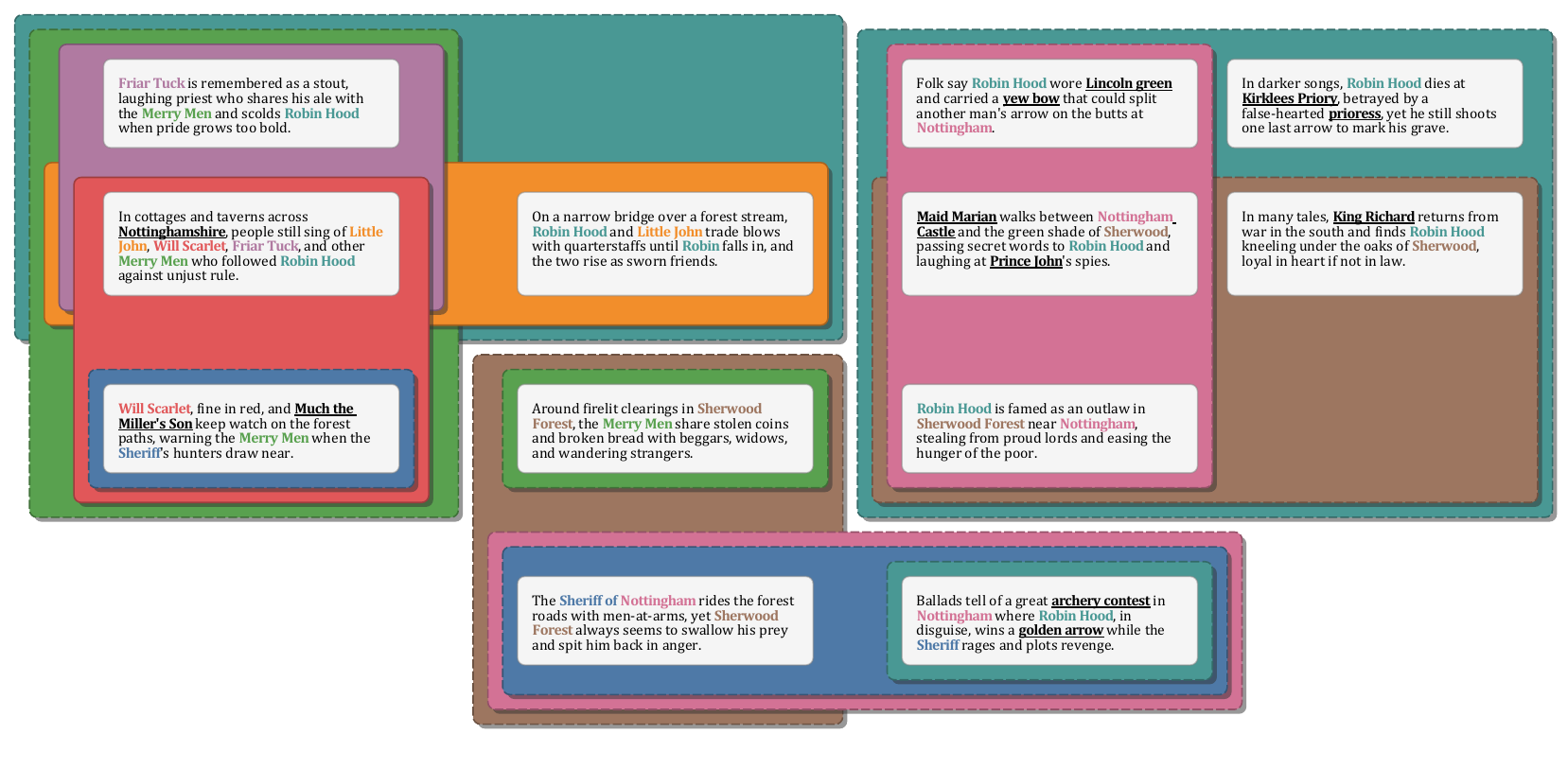}
        \caption{BlockSets with rectangles for the Robin Hood tales.}
        \label{fig:robin-hood:rectangles}
\end{figure}

\mypar{Different shape types.}
Lastly, we compare BlockSets using different shape types.
We do not consider arbitrary polygons as orthoconvex polygons in detail are often already maximally compact (\autoref{fig:teaser}), and in other cases, arbitrary polygons would lead to visualizations that are hard to read (see \autoref{fig:murder:MosaicSets} and the \autoref{fig:murder:BlockSets-render:MosaicSets} in the supplementary material).
In Figures~\ref{fig:robin-hood:nabla}--\ref{fig:robin-hood:rectangles} we see that as the shape type becomes less flexible, the visualization becomes less compact but the shapes become more structured.
When using rectangles, the input needs to be split into components to create a compact visualization.
Which shape type works best differs per dataset.
For the Robin Hood tales, the orthoconvex shapes of \autoref{fig:teaser} create a compact and readable visualization; using more structured shapes makes the visualization considerably less compact.
However, for \autoref{fig:EU:BlockSets} and \autoref{fig:murder:BlockSets} $\nabla$-shapes provide a compact visualization and are the preferred option.

\begin{figure}
        \includegraphics[width=\linewidth]{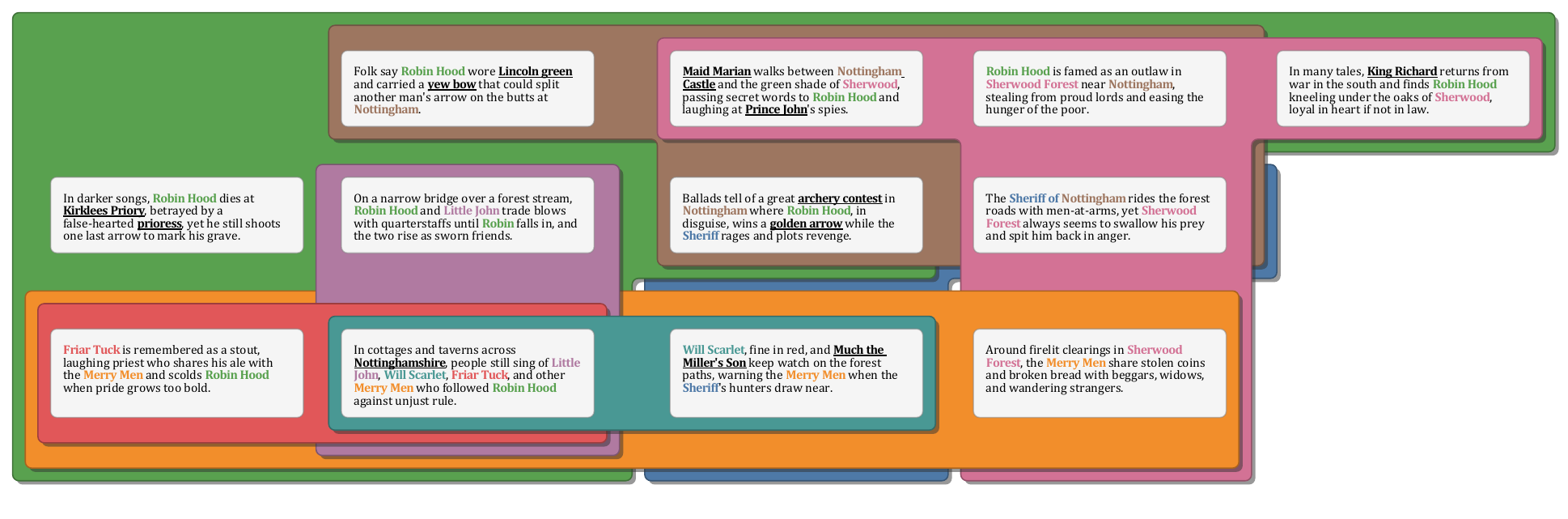}
        \caption{BlockSets with $\nabla$-shapes for the Robin Hood tales.}
        \label{fig:robin-hood:nabla}
\end{figure}

\section{Future Work}
BlockSets are a novel structured approach to visualizing set systems with large data. Our design decisions and algorithmic solutions are guided by specific quality criteria and requirements, which are rooted in the literature and imposed by the data types we aim to visualize. We believe that the resulting visualizations can indeed serve users well when interacting with large set data and, specifically, with statement-entity data. We found that our colleagues were easily able to solve the murder mystery and other related puzzles when presented with them in BlockSets form. However, it would be beneficial to explore the usability of BlockSets, also for other data types, with a larger group in a proper user study.

On the algorithmic side, we would like to further explore the different options for splitting algorithms. As mentioned before, RectEuler takes an element-centric approach, while we prefer a set-centric method. However, in general, the best approach likely depends on the type of set data to be visualized. We would hence like to further explore which characteristics of the data are served best by which type of splitting and develop automated (hybrid) methods that can choose the best split for the data at hand.

Finally, we sketched in Figures~\ref{fig:EU} and~\ref{fig:small-multiple} how to possibly augment BlockSets with labels, in case that the element data does not carry the set information directly. Developing the corresponding automated labeling methods is an interesting algorithmic challenge.

\mypar{Acknowledgements.}
We thank Tomasz Soróbka for helpful discussions on initial designs.


\clearpage
\printbibliography

\clearpage

\pagestyle{empty}

\begin{figure*}[t]
\begin{minipage}{\textwidth}
\centering
{\bfseries\huge%
BlockSets: A Structured Visualization for Sets with Large Elements
\vskip 5pt
\bfseries\LARGE Supplementary Material\\}
\vskip 25pt
{\normalsize%
Neda Novakova, Veselin Todorov, Steven van den Broek, Tim Dwyer, and  Bettina Speckmann\\
\vspace{25pt}
}
\end{minipage}
\end{figure*}

\setcounter{section}{0}

\section{Additional figures.}

\begin{figure*}[b]
    \centering
    \includegraphics[width=\linewidth]{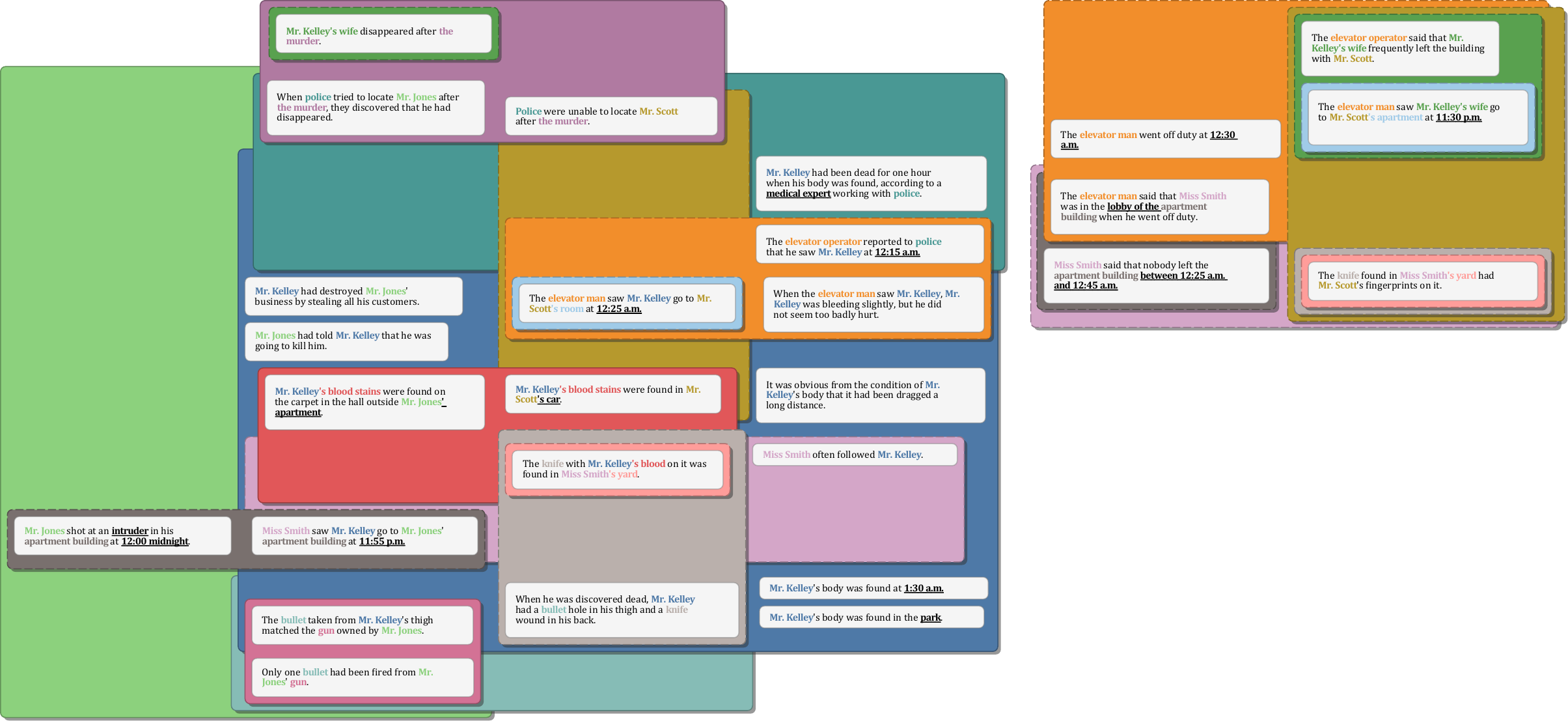}
    \caption{RectEuler~\cite{paetzold2023recteuler} solution of the murder mystery from \autoref{fig:murder} rendered in BlockSets style. To create this rendering, we adapted the ILP of RectEuler to not use set labels or set circle markers, and to not group statements that belong to the same set of sets.}
    \label{fig:murder:BlockSets-render:RectEuler}
\end{figure*}

\begin{figure*}
    \centering
    \includegraphics[width=\linewidth]{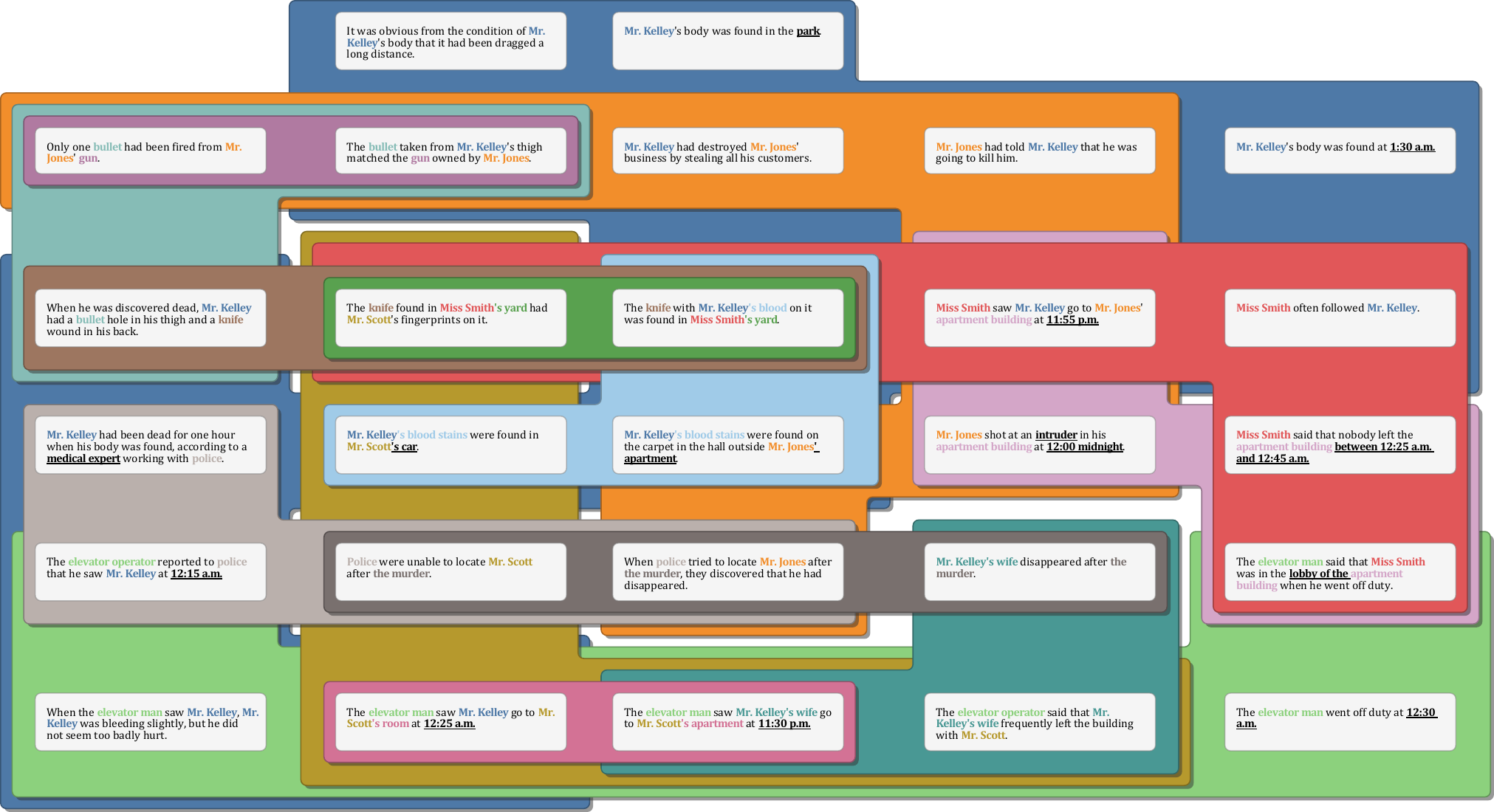}
    \caption{MosaicSets~\cite{DBLP:journals/tvcg/RottmannWBGNH23} solution of the murder mystery from \autoref{fig:murder} rendered in BlockSets style.}
    \label{fig:murder:BlockSets-render:MosaicSets}
\end{figure*}

\begin{figure*}
    \includegraphics[width=\linewidth]{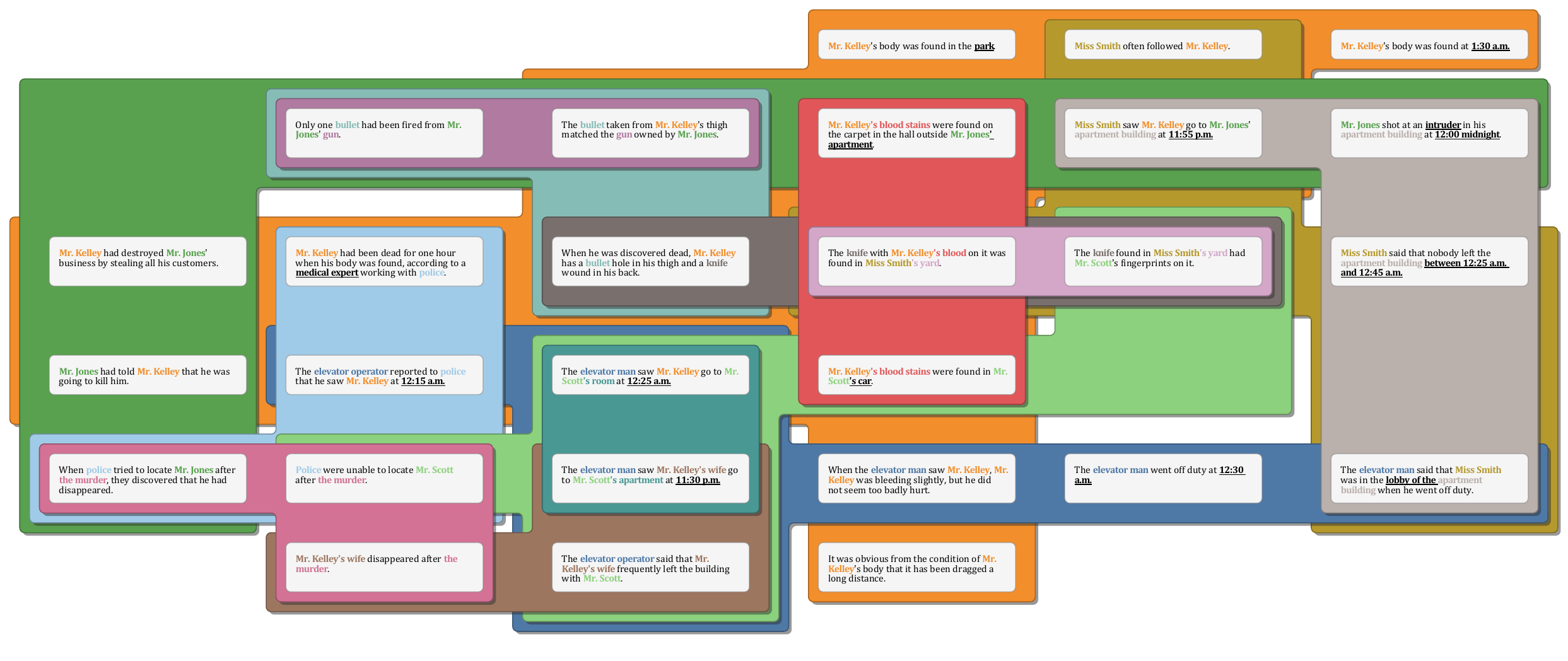}
    \caption{BlockSets solution of the murder mystery from \autoref{fig:murder} with orthoconvex shapes.}
    \label{fig:murder:BlockSeys:orthoconvex}
\end{figure*}

\begin{figure*}
    \begin{subfigure}[t]{\linewidth}
        \includegraphics[width=\textwidth]{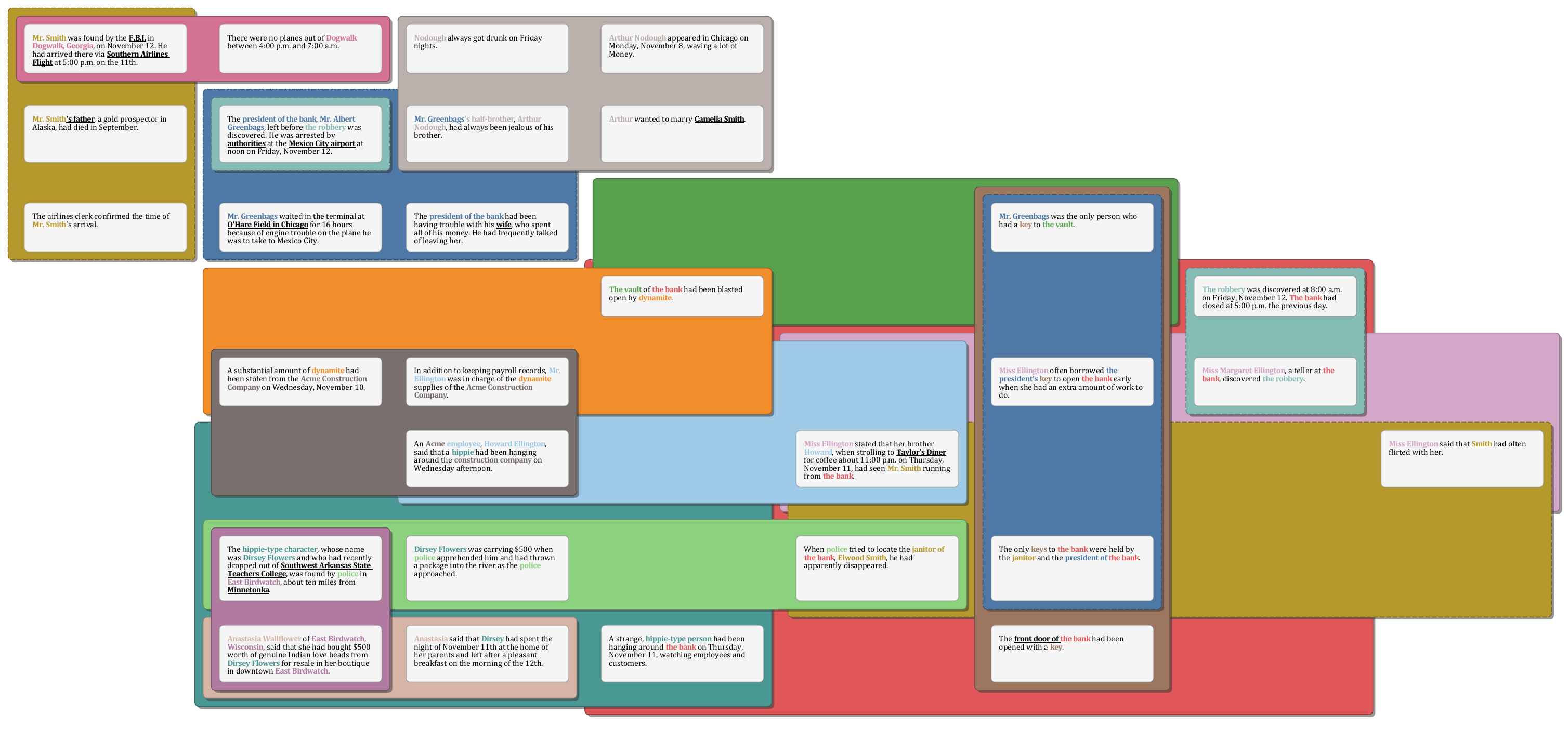}
        \caption{BlockSets with rectangles. Only two sets are duplicated.}
        \label{fig:bank-robbery:BlockSets}
    \end{subfigure}

    \vspace{5mm}
    
    \begin{subfigure}{\linewidth}
        \centering
        \includegraphics[width = \linewidth]{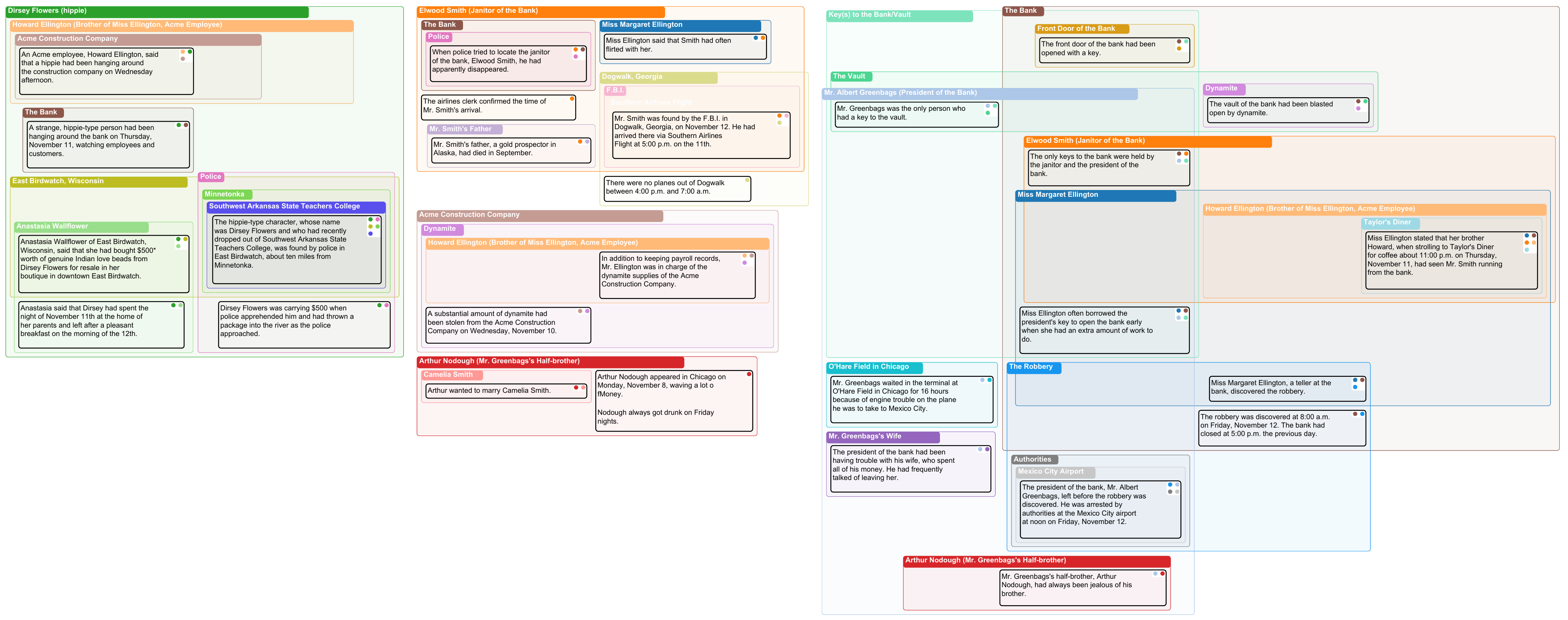}
        \caption{RectEuler~\cite{paetzold2023recteuler} with one minute time limit. Eight sets are duplicated, two of which are duplicated thrice.}
        \label{fig:bank-robbery:RectEuler:split}
    \end{subfigure}
    \caption{Comparison of splitting algorithm of BlockSets and RectEuler on a bank robbery mystery puzzle~\cite{stanford1969learning}. See \autoref{fig:bank-robbery:RectEuler:no-split} for a version of RectEuler that does not split.}
    \label{fig:bank-robbery}
\end{figure*}

\begin{figure*}
    \includegraphics[width = \linewidth]{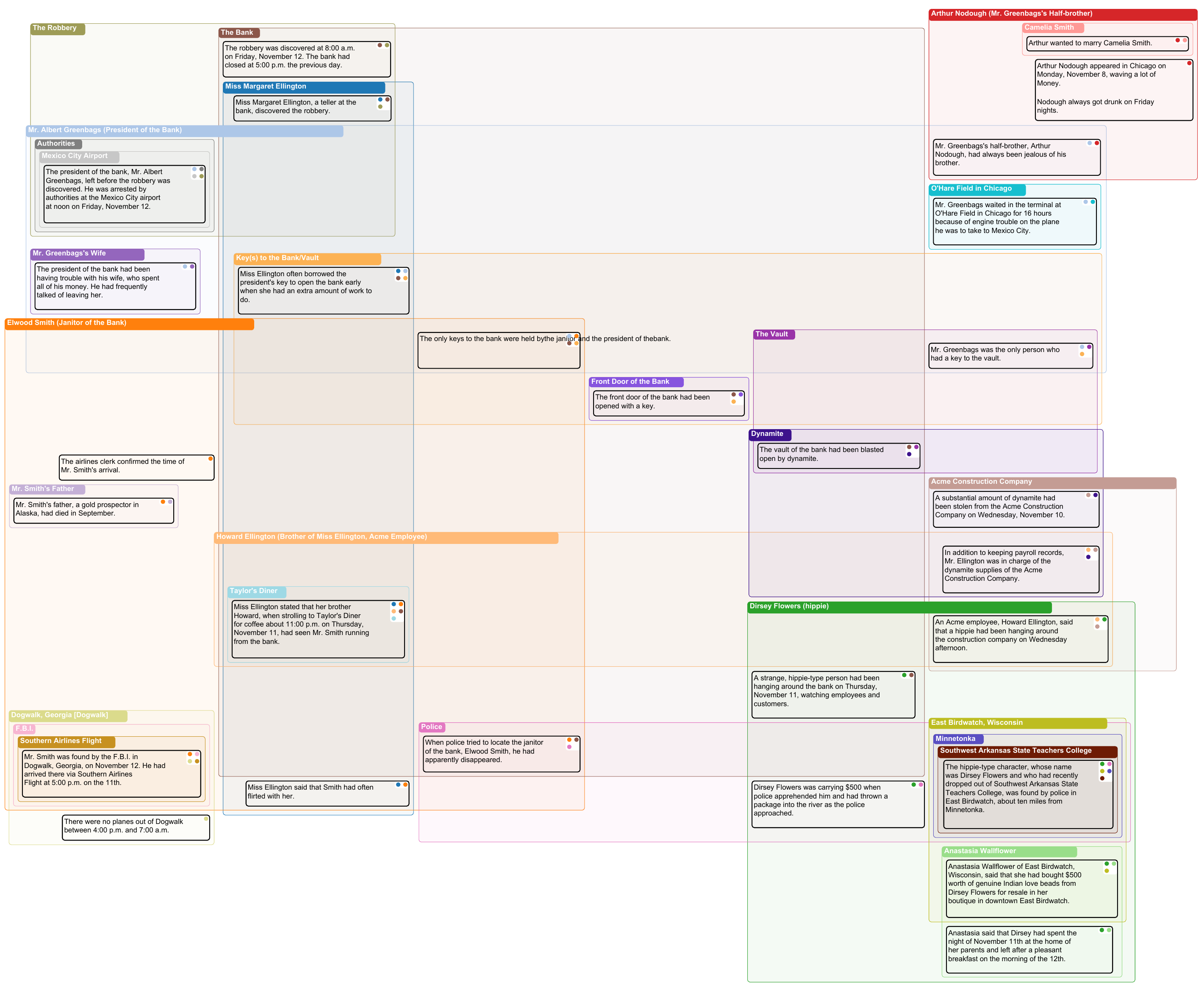}
    \caption{RectEuler~\cite{paetzold2023recteuler} on the bank robbery mystery puzzle with 100 minute time limit. RectEuler splits only when it cannot find a feasible solution, BlockSets also splits when solutions are not sufficiently compact.}
    \label{fig:bank-robbery:RectEuler:no-split}
\end{figure*}
\end{document}